\documentclass[preprint,nofootinbib,floatfix,a4paper,prd,superscriptaddress]{revtex4-1}   	
\usepackage{geometry}                		
\usepackage[colorlinks,citecolor=blue]{hyperref}
\usepackage{amsmath}
\usepackage{mathrsfs}
\usepackage{bbm}
\usepackage{amsfonts}
\usepackage{amssymb}
\usepackage{latexsym}
\usepackage{graphicx}
\usepackage[english]{babel}
\usepackage{multirow}
\usepackage{float}
\usepackage{url}
\usepackage{slashed}
\usepackage{ulem}

\usepackage{orcidlink}

\newcommand{\be}{\begin{equation}}
\newcommand{\ee}{\end{equation}}
\newcommand{\ba}{\begin{array}}
\newcommand{\ea}{\end{array}}
\newcommand{\bea}{\begin{eqnarray}}
\newcommand{\eea}{\end{eqnarray}}

\usepackage[utf8]{inputenc}
\usepackage[T1]{fontenc}

\newcommand{\PU}{\affiliation{\small Phenikaa Institute for Advanced Study, Phenikaa University, Nguyen Trac, Duong Noi, Hanoi 100000, Vietnam}}

\newcommand{\NCTS}{\affiliation{\small Physics Division, National Center for Theoretical Sciences, National Taiwan University, Taipei 106319, Taiwan}}

\newcommand{\NTNU}{\affiliation{\small Department of Physics, National Taiwan Normal University, Taipei 116, Taiwan}}

\begin{document}

\title{Search for Long-Lived Dark Photons from Dark Radiation at the LHC}
\author{Chuan-Ren Chen \orcidlink{0000-0002-3220-2550}} 
\email{crchen@ntnu.edu.tw} \NTNU

\author{Van Que Tran \orcidlink{0000-0003-4643-4050}}
\email{vqtran@phys.ncts.ntu.edu.tw} \NCTS \PU

\begin{abstract}
We investigate a novel production mechanism for long-lived dark photons at the LHC, arising from dark radiation emitted from $\chi$ in $Z\to\bar{\chi}\chi$ decays, where $\chi$ is a fermionic dark matter candidate.  
The effective $Z\chi\chi$ coupling is generated radiatively through one-loop diagrams involving the top quark and a new colored scalar.  
We show that dark photons produced via this dark radiation channel can dominate over the conventional sources---meson decays and proton bremsstrahlung---across wide regions of parameter space, particularly for small kinetic mixing and dark photon masses well above the GeV scale.  
Using this enhanced production mechanism, we analyze the sensitivity of dedicated long-lived particle detectors, including FASER2, FACET, and MATHUSLA. 
We find that these experiments can significantly surpass existing bounds, probing regions of dark photon parameter space consistent with the observed dark matter relic abundance and inaccessible in conventional dark photon scenarios.
\end{abstract}

\maketitle

\section{Introduction}

Astrophysical and cosmological observations provide compelling evidence for the existence of dark matter (DM), which dominates the matter density of the Universe. However, the particle nature of DM remains unknown. The absence of signals in direct, indirect, and collider searches indicates that the DM particle may interact extremely weakly with the Standard Model (SM) sector, for example through loop effects or feeble portal interactions.  
 
A minimal and well-motivated realization consists of a Dirac fermion DM candidate $\chi$ residing in a dark sector charged under a broken $U(1)$ gauge symmetry, which communicates with the SM via kinetic mixing~\cite{Pospelov:2007mp}. The corresponding massive gauge boson, the dark photon $A'$, mediates the interactions between the two sectors. This framework accommodates the observed relic abundance over a wide range of parameter space. 
It should be noted that, for $m_{A'} < m_\chi$, efficient annihilation $\bar{\chi}\chi \to A'A'$ typically drives the DM abundance below the observed value unless the dark gauge coupling is sufficiently suppressed. Conversely, when $m_{A'} \gtrsim m_\chi$, the relic density can be set by the kinematically forbidden annihilation channel which is exponentially sensitive to the mass splitting between the dark photon and DM~\cite{DAgnolo:2015ujb}.

At the Large Hadron Collider (LHC), dark photons can be probed through meson decays, proton bremsstrahlung, and direct production via $pp \to A'$, followed by visible decays into SM fermions~\cite{LHCb:2019vmc,LHCb:2020ysn,CMS:2019buh,CMS:2023hwl,FASER:2023tle}. In these conventional searches, both the production rate and decay width scale as $\epsilon^2$, where $\epsilon$ denotes the kinetic mixing parameter. As a result, resonance searches are primarily sensitive to heavier dark photons with relatively large $\epsilon$, while smaller values of $\epsilon$ suppress production and lead to displaced or even detector-escaping decays. Dedicated long-lived particle (LLP) detectors such as FASER, FASER2, FACET, and MATHUSLA instead target light dark photons ($\lesssim \mathrm{GeV}$) produced via meson decays and bremsstrahlung, within the limited range of $\epsilon$ that yields observable decay lengths.

In this work, we propose a complementary production mechanism based on dark radiation emitted in DM production processes~\cite{Buschmann:2015awa, Du:2021cmt,Du:2019mlc}. In particular, we consider DM pairs produced via on-shell $Z$ boson decays, followed by final-state radiation of dark photons. We denote this new dark photon production channel as dark radiation. 
A key feature of this new mechanism is the parametric decoupling between production and decay: the production rate is governed by ${\rm BR}(Z \to \bar{\chi}\chi)$ and the dark radiation probability, while the visible decay of $A'$ depends solely on $\epsilon$. This is in sharp contrast to conventional setups, where both production and decay are controlled by the same small parameter.

Owing to the enormous number of $Z$ bosons produced at LHC ($\sigma(pp \to Z) \approx 60~{\rm nb}$ at $\sqrt{s}=14~{\rm TeV}$)~\cite{ATLAS:2016fij,ATLAS:2024irg,CMS:2025sgv}, the dark radiation channel can dominate over the meson decays and proton bremsstrahlung across substantial regions of parameter space. Consequently, the sensitivity of LLP detectors to dark photons with small kinetic mixing and higher masses is significantly enhanced.
We analyze the reach of FASER2, FACET, and MATHUSLA, and demonstrate that they can significantly surpass current experimental bounds on long-lived dark photons (LLDPs) and probe portions of parameter space inaccessible to the conventional dark photon model.

This paper is organized as follows. In Sec.~\ref{sec:model} we introduce the model and derive the effective interactions between DM and $Z$ boson relevant for DM annihilation, direct detection, and collider production. 
We also present a detailed analysis of the loop-induced coupling between the $Z$ boson and the DM pair.
Sec.~\ref{sec:exp_constraint} summarizes the existing experimental constraints, including those from $Z$ invisible decay, direct-detection null results for the search of DM, monojet searches at the LHC and the cosmic microwave background (CMB) observations. In Sec.~\ref{sec:dmrelic} we analyze the thermal relic abundance, including the contributions from forbidden and s-channel annihilation channels. Sec.~\ref{sec:llp} presents the analysis of LLDPs produced from dark radiation at FASER2, FACET, and MATHUSLA. We show the numerical results in Sec.~\ref{sec:result} and conclude in Sec.~\ref{sec:conclusion}. Appendices show details of the dark sector model, including constraints and expressions of form factors.

\section{The Model}
\label{sec:model}

We consider a dark sector charged under a dark Abelian gauge group $U(1)_D$.
The dark sector contains a vector-like Dirac fermion, $\chi$, with unit $U(1)_D$ charge, which can serve as a DM candidate. The corresponding gauge boson, the dark photon $A'_\mu$, mixes kinetically with the SM hypercharge gauge boson, thereby inducing feeble interactions between the dark photon and SM fermions.
This represents the minimal and well-studied portal between the visible and dark sectors.

To link the dark fermion to SM matter, we extend the minimal model with a colored scalar mediator $\Phi$, transforming in the fundamental representation of $SU(3)_C$, singlet under $SU(2)_L$, and carrying both hypercharge and dark charge.
With these quantum numbers, $\Phi$ can couple the dark fermion $\chi$ to the SM right-handed quarks $q_R$, providing an additional portal between the two sectors. For illustration, we consider the case in which the SM right-handed top quark, $t_R$, couples to $\chi$ and $\Phi$. Gauge invariance under $U(1)_Y$ then fixes the hypercharge of $\Phi$ to be $Y_{\Phi} = Y_{t_R} = 2/3$.
The phenomenological role of $\Phi$ is analogous to squarks in supersymmetric theories. 
The particle content and quantum numbers of the dark sector in our setup are summarized in Table~\ref{tab:particles}.
\begin{table}[t]
\centering
\begin{tabular}{|c|c|c|c|}
\hline
Field & Spin & $(SU(3)_c,SU(2)_L,U(1)_Y)$  & $U(1)_D$  \\
\hline
$\chi$ & $1/2$ & $(\mathbf{1},\mathbf{1},0)$ & $+1$  \\
$\Phi$ & $0$ & $(\mathbf{3},\mathbf{1},Y_\Phi)$ & $-1$  \\
$A'_\mu$ & $1$ & $(\mathbf{1},\mathbf{1},0)$ & --  \\
\hline
\end{tabular}
\caption{Particle content and charge assignments in the model. $Y_\Phi$ is hypercharge of $\Phi$.}
\label{tab:particles}
\end{table}

The relevant interaction terms in the Lagrangian are given by 
\bea
\label{eq:Lag}
\mathcal{L} &\supset& 
- \frac{1}{4} F'_{\mu\nu} F'^{\mu\nu} 
- \frac{\epsilon}{2 c_W} F'_{\mu\nu} F^{\mu\nu} 
+ (D_\mu \Phi)^\dagger (D^\mu \Phi) 
+ \bar{\chi} (i \slashed{D} - m_\chi)\chi  \nonumber \\
&& - \,\frac{1}{2} m_{A'}^2 A'_\mu A^{\prime\mu}  - \, \frac{1}{2}m_{\Phi}^2 |\Phi|^2 +\, \lambda_{H\Phi} |H|^2 |\Phi|^2  
+ \lambda_{\chi t} \, \Phi^\ast \, \bar{\chi} \, t_R + \text{h.c.}\, ,\, ,
\eea
where $c_W$ stands for cosine of the weak mixing angle, $D_\mu$ denotes the covariant derivative including both SM and $U(1)_D$ gauge interactions, $F_{\mu\nu}$ ($F'_{\mu\nu}$) is the field strength tensor of the hypercharge (the dark photon), $H$ is the SM $SU(2)_L$ doublet Higgs and $\epsilon$ parametrizes the strength of the kinetic mixing. 

Although $\Phi$ can be pair-produced at the LHC via QCD interactions, existing searches for top-quark pairs accompanied by missing transverse energy (MET),
($pp \to \Phi \Phi^* \to t\bar{t}\chi\bar{\chi}$),
already impose stringent lower bounds on its mass. In particular, for a light $\chi$, scalar mediator masses up to $m_\Phi \simeq 1.2~\mathrm{TeV}$ are excluded under the assumption of decay branching ratio $\mathrm{BR}(\Phi \to t \chi)=100\%$~\cite{CMS:2021beq,ATLAS:2024rcx}.
These bounds are significantly weakened in the compressed regime, $m_\Phi \simeq m_\chi + m_t$, where current limits extend down to $m_\Phi \sim 780~\mathrm{GeV}$~\cite{CMS:2025ttk}.
Importantly, for $m_\Phi \gtrsim \mathcal{O}(\mathrm{TeV})$, the pair-production cross section falls below the femtobarn level~\cite{Beenakker:2024jwh}, rendering this production mode phenomenologically subdominant for the LLDP signals considered in this work.

In addition, the trilinear Higgs-portal interaction $h \Phi \Phi^*$, where $h$ denotes the 125~GeV Higgs boson, arises from the quartic operator $|H|^2 |\Phi|^2$ in Eq.~(\ref{eq:Lag}). This coupling induces loop-level corrections to Higgs production and decay processes. Consequently, precision measurements of Higgs signal strengths constrain the portal coupling $\lambda_{H\Phi}$, stringent for colored scalar masses below the TeV scale~\cite{CMS:2018uag, ATLAS:2022vkf}. Details of the relevant collider bounds are provided in Appendix~\ref{app:constraints_colored_scalar}.

Even in the regime where $\Phi$ is too heavy to be efficiently produced at the LHC, its Yukawa interaction with $\chi$ and the SM right-handed top quark, as given in Eq.~(\ref{eq:Lag}), induces effective couplings between electroweak gauge bosons and the dark fermion at the loop level. In particular, the triangle diagram involving $(t_R, \Phi, \chi)$ generates effective $h\bar{\chi}\chi$ and $Z\bar{\chi}\chi$ interactions.

While the Higgs-mediated channel is in principle present, its phenomenological impact is strongly suppressed. The inclusive Higgs production cross section at the LHC is relatively small, $\mathcal{O}(60~\mathrm{pb})$~\cite{ATLAS:2023tnc}, and is further constrained by stringent bounds on the Higgs invisible decay width. As a result, it does not yield an observable rate of LLDPs from dark radiation in the scenario considered here.
In contrast, the large $Z$ boson yield at the LHC, combined with the loop-induced $Z \to \bar{\chi}\chi$ decay, can lead to sizable production rates\footnote{The $Z$ portal to the dark sector has been studied in various contexts~\cite {Liu:2017zdh, Cheng:2019yai, Cheng:2024hvq}.}. We therefore focus exclusively on the $Z$ boson channel in the following analysis.

\begin{figure}[t]
\centering
\includegraphics[width=0.49\textwidth]{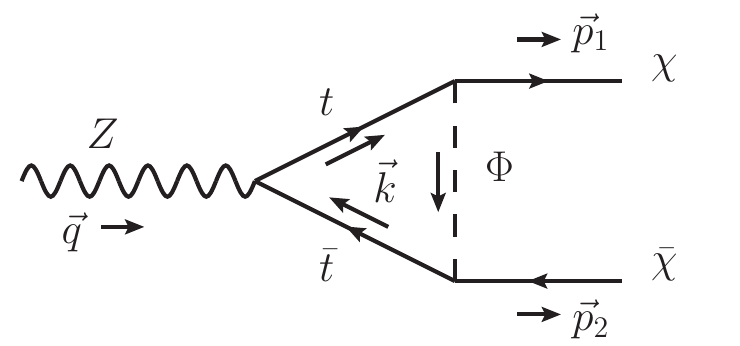}
\includegraphics[width=0.49\textwidth]{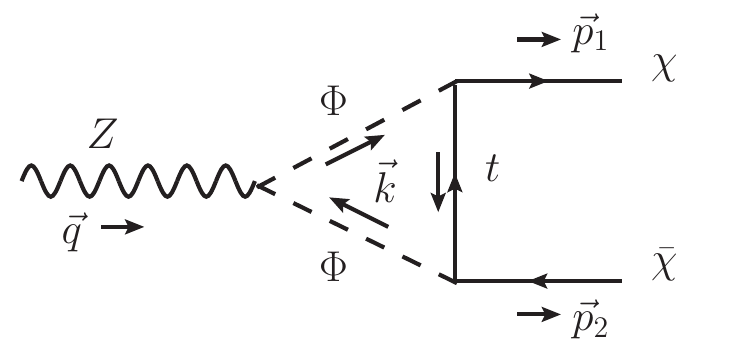}
\caption{One-loop triangle diagrams contributing to $Z \to \chi \bar{\chi}$.}
\label{fig:Zchichi_loop}
\end{figure}

An effective $Z\bar{\chi}\chi$ interaction can be radiatively generated at one-loop level via the exchange of the top quark and the colored scalar $\Phi$, as shown in Fig.~\ref{fig:Zchichi_loop}. 
The resulting effective vertex can be parameterized in terms of Lorentz-invariant form factors as
\begin{align}
\Gamma^\mu(q^2)
= \gamma^\mu \left[ F_L(q^2) P_L + F_R(q^2) P_R \right]
+ \frac{i\sigma^{\mu\nu}q_\nu}{2m_\chi}
\left[ F_M(q^2) + F_E(q^2)\gamma^5 \right],
\end{align}
where $q$ is the momentum carried by the $Z$ boson, $P_{R,L} = \frac{1}{2} (1\pm \gamma_5)$, $\sigma^{\mu\nu} = \frac{i}{2} [\gamma^\mu, \gamma^\nu] = \frac{i}{2} (\gamma^\mu\gamma^\nu - \gamma^\nu \gamma^\mu)$ and $F_{L, R, M, E}$ is the form factor. Analytical expressions of the form factors are given in Appendix~\ref{app:formfactors}. 
And we have confirmed that the ultraviolet divergences are exactly canceled between these two classes of diagrams.
This cancellation can be traced back to the relation between the $Z$ couplings of the top quark and the scalar mediator which ensures gauge invariance of the effective interaction.

Phenomenologically, we find that the interaction is dominated by the left-handed vector form factor $F_L$, while the right-handed and dipole contributions ($F_R$, $F_M$) are subleading in the parameter space of interest as shown in the left panel of Fig.~\ref{fig:Zloop} in the Appendix~\ref{app:formfactors}. The electric dipole form factor vanishes identically, $F_E = 0$, reflecting the absence of CP-violating phases in the Yukawa sector. Therefore, we approximately obtain the 
effective Lagrangian for the $Z\bar{\chi}\chi$ interaction as 
\be
\label{eq:Zeff}
 \mathcal{L}_{\rm eff} \simeq g_Z^\chi \, \bar{\chi} \gamma^\mu P_L \chi \, Z_\mu, \quad {\rm where} \quad g_Z^\chi \equiv F_L(q^2 = m_Z^2). 
\ee

\begin{figure}[htbp!]
\centering
\includegraphics[width=0.65\textwidth]{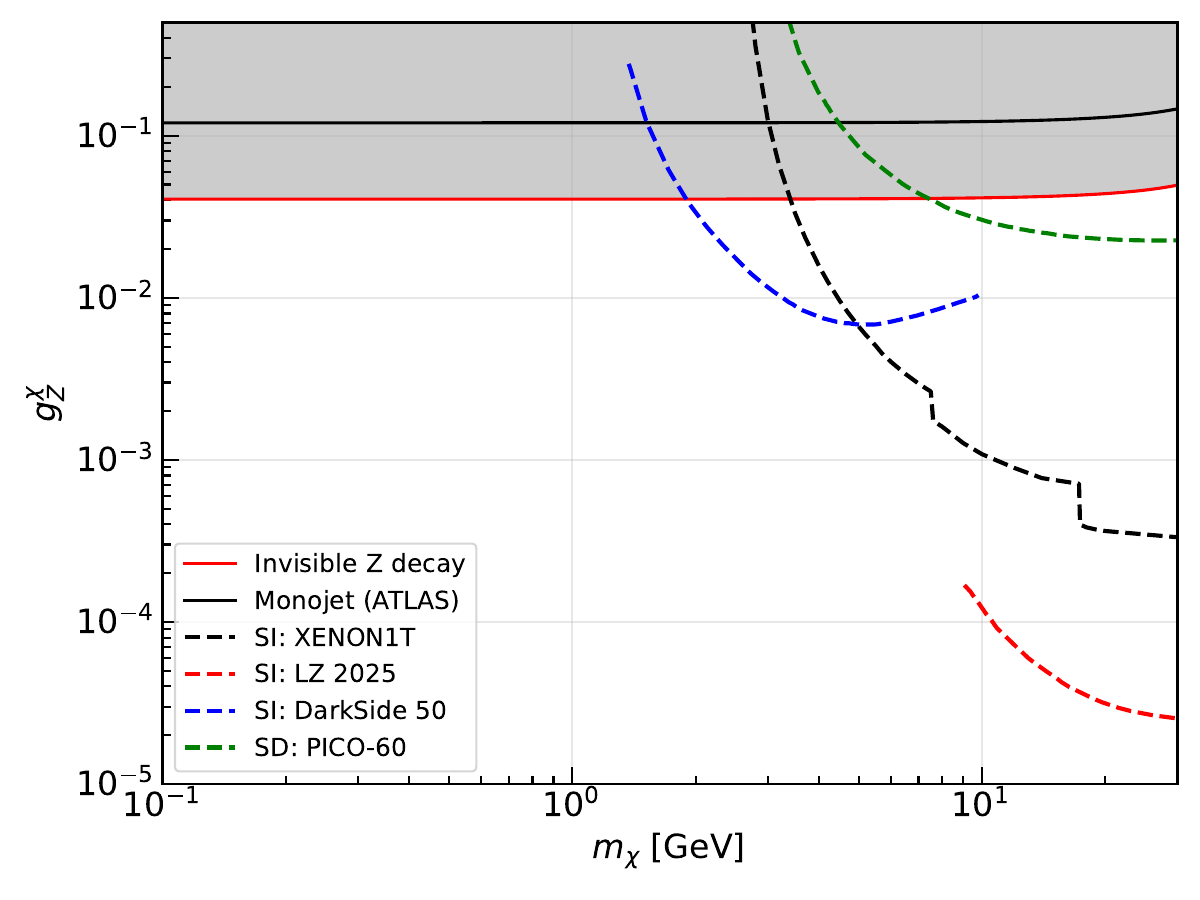}
\caption{\label{fig:constraint_mchi_gZchi} The exclusion limits from invisible $Z$ decay (solid red), ATLAS monojet (solid black) and DM direct detections (dashed curves) projected on ($m_\chi, g_Z^\chi$) plane.}
\end{figure}

\section{Experimental Constraints}
\label{sec:exp_constraint}
In this section, we discuss the current experimental constraints on the loop-induced $Z\bar{\chi}\chi$ interaction. The relevant constraints include invisible $Z$ decay, monojet searches at ATLAS, DM direct detections and CMB observations. The results are depicted in Fig.~\ref{fig:constraint_mchi_gZchi}.

\paragraph{Invisible $Z$ decay:}
If kinematically allowed, the $Z$ boson can decay into a pair of dark fermions $\chi$ via the loop-induced coupling in Eq.~(\ref{eq:Zeff}).
The partial width is given by
\be
\label{eq:Ztochichi}
\Gamma(Z\to  \bar{\chi} \chi) \simeq \frac{\left|g_Z^\chi \right|^2}{24 \pi} m_Z \left( 1 - \frac{m_\chi^2}{m_Z^2} \right) \left( 1- \frac{4 m_\chi^2}{m_Z^2}\right)^{1/2}.
\ee 
LEP measures the invisible $Z$ width to precision $\Delta\Gamma_Z^{\rm inv} < 2~{\rm MeV}$ \cite{ALEPH:2005ab},
which  corresponds to ${\rm BR}(Z\rightarrow\chi\chi) \lesssim 8\times 10^{-4}$ given the $Z$ total width is $2.4955$ GeV~\cite{ParticleDataGroup:2024cfk}. 
Using Eq.~(\ref{eq:Ztochichi}), this places the most stringent bound on $g_Z^\chi$ for the DM mass $m_\chi \lesssim 2$ GeV as shown by the red curve in Fig.~\ref{fig:constraint_mchi_gZchi}.

\paragraph{Dark Matter Direct Detection:}
The loop-induced coupling in Eq.~(\ref{eq:Zeff}) contains both vector and axial-vector which can give rise to 
spin-independent (SI) and spin-dependent (SD) DM--nucleon scattering via $Z$ boson exchange,
respectively~\footnote{The dark photon and loop-induced Higgs exchange also contribute to the DM--nucleon scattering cross section, but these effects are negligible in the parameter region of interest.}. 

At momentum transfer $q^2 \ll m_Z^2$, the corresponding cross sections are given by \cite{Abdallah:2015ter}
\be
\label{eq:DMDD}
\sigma_{\chi N}^{\rm SI}
=\frac{\mu_{\chi N}^2}{\pi m_Z^4} \left| f_N\right|^2 , \;\;\;\;\;\;\;\; \sigma_{\chi N}^{\rm SD}, 
=\frac{3 \mu_{\chi N}^2}{\pi m_Z^4} \left|a_N\right|^2
\ee
where $N$ denotes either proton $p$ or neutron $n$, and $\mu_{\chi N}$ = $m_\chi m_N/(m_\chi + m_N)$. The couplings $f_N$ and $a_N$ stand for the effective nucleon couplings which are given as 
\be
f_p = \frac{g_Z^\chi}{2}\left( 2 g_u^V + g_d^V \right) , \;\;\;\;\;\;\;\; f_n = \frac{g_Z^\chi}{2}\left( g_u^V + 2 g_d^V \right), 
\ee
and 
\be
a_{p,n} = \frac{g_Z^\chi}{2} \sum_{q = u,d,s} \Delta q^{(p,n)} g_q^A 
\ee
with $g_q^{V}$ and $g_q^{A}$ are the vector and axial vector coupling of $Z$ to SM quarks, respectively. 
The coefficients $\Delta q^{(N)}$ encode the contributions of the light quarks to the nucleon spin which are given by~\cite{ParticleDataGroup:2014cgo}
\bea
\Delta u^{(p)} &=& \Delta d^{(n)} = 0.84 \pm 0.02, \nonumber \\
\Delta d^{(p)} &=& \Delta u^{(n)} = -0.43 \pm 0.02,  \\
\Delta s^{(p)} &=& \Delta s^{(n)} = -0.09 \pm 0.02. \nonumber 
\eea

We note that spin-independent DM--nucleon scattering via $Z$ exchange is
isospin violating (ISV), i.e.\ $f_p \neq f_n$. Since most experimental
limits on the SI cross section are reported assuming isospin conservation
($f_p = f_n$), the published bounds must be rescaled as
 $\sigma_{\rm limit} \;\to\; \sigma_{\rm limit}\, F_Z $,
to correctly account for ISV effects. The rescaling factor $F_Z$ is given by
\cite{Feng:2011vu}
\be
F_Z \;=\;
\frac{\displaystyle\sum_i \eta_i\, \mu_{A_i}^2\, A_i^2}
{\displaystyle\sum_i \eta_i\, \mu_{A_i}^2\,
\left[ Z_{\rm atom} + (A_i - Z_{\rm atom})\, f_n/f_p \right]^2},
\ee
where $\mu_{A_i} = m_\chi m_{A_i}/(m_\chi + m_{A_i})$ is the reduced
DM--nucleus mass, $Z_{\rm atom}$ is the atomic number, and $\eta_i$ and $A_i$
denote the natural abundance and mass number of the $i$$^{\rm th}$ isotope, respectively.

The resulting limits from LZ~\cite{LZ:2024zvo}, XENON1T~\cite{XENON:2019gfn},
DarkSide-50~\cite{DarkSide-50:2022qzh,DarkSide-50:2025lns}---after applying the ISV correction---are shown as
dashed curves in the $(m_\chi, g_Z^\chi)$ plane in
Fig.~\ref{fig:constraint_mchi_gZchi}. The SD cross section limit from PICO-60~\cite{PICO:2019vsc} is also included (dashed green line), however it is weaker than the SI limits. These experiments impose the strongest
constraints for $m_\chi \gtrsim 2$~GeV, forcing the effective coupling
$g_Z^\chi$ to be very small and correspondingly suppressing the branching ratio
${\rm BR}(Z\to\chi\bar{\chi})$. As a result, the $Z$-mediated production of
$\chi$ pairs at the LHC is strongly limited in this mass regime. However, if
$\chi$ constitutes only a subcomponent of the total DM
abundance, the direct detection bounds are weakened accordingly, allowing a
larger ${\rm BR}(Z\to\chi\bar{\chi})$ and enabling a sizeable production rate.

\paragraph{Monojet constraints: }
Searches at the LHC for events featuring MET 
accompanied by one or more energetic jets impose stringent bounds on DM interactions mediated by vector/scalar bosons~\cite{CMS:2021far, ATLAS:2021kxv}.
In our setup, monojet constraints are dominated by the $Z$-exchange contribution.

When $2 m_\chi \leq m_Z$, the production rate can be approximately given as 
\begin{equation}
\sigma(pp \to \chi\bar{\chi} + j)
\simeq \sigma(pp \to Z + j) \times  {\rm BR}(Z \to\bar{\chi}  \chi).
\end{equation}

Since $Z \to\bar{\chi}  \chi$ produces MET and jet kinematics nearly identical to $Z\to\nu\bar\nu$ for light $\chi$, the detector acceptance and reconstruction efficiencies are well-approximated by that of the SM process. Using the data from ATLAS~\cite{ ATLAS:2021kxv}, 
we found the optimal selection region is to be in 
the window: MET $\in (300,350)$ GeV 
(the EM2 region in Ref.\ \cite{ ATLAS:2021kxv}) which yields a constraint ${\rm BR}(Z \to \chi\bar{\chi})  <  7\times10^{-3}$. 
This is about one order of magnitude weaker than the $Z$ invisible decay constraint discussed above.

\paragraph{CMB constraints:}
The annihilation channel $\chi\bar{\chi}\to f\bar f$ via $Z$-boson exchange proceeds through an $s$-wave process and therefore remains efficient at late times. 
Consequently, it is constrained by observations of CMB, which are sensitive to energy injection from DM annihilation during the recombination epoch.

Taking the most stringent limits from Planck collaboration~\cite{Planck:2018vyg}, which arise from annihilation into electron--positron pairs, we obtain the approximate bound
\begin{equation}
\langle\sigma v\rangle_{\rm CMB}
\;\lesssim\;
6\times10^{-28}\left(\frac{m_\chi}{1~\mathrm{GeV}}\right)
{\rm cm^3\,s^{-1}}.
\end{equation}

In this model, the thermally averaged annihilation cross section of $\chi\bar{\chi}$ into SM fermions via $Z$-boson exchange is given by
\begin{equation}
\label{eq:chichiff}
\langle \sigma v\rangle_{\chi \bar{\chi} \to \bar{f} f} \simeq 
\frac{N_f \left(|g^V_f|^2 + |g^A_f|^2 \right)  \left|g_Z^\chi/2 \right|^2 }{2\pi} 
\frac{2 m_\chi^2 + m_f^2}{\left(4m_\chi^2 - m_Z^2\right)^2 + m_Z^2\Gamma_Z^2} 
\sqrt{1- \frac{m_f^2}{m_\chi^2}},
\end{equation}
where $N_f$ is the color factor, and $g_f^{V}$ and $g_f^{A}$ denote the vector and axial-vector couplings of the $Z$ boson to SM fermions, respectively.

Away from the $Z$-pole region ($m_\chi \ll m_Z/2$), imposing the CMB bound on Eq.~(\ref{eq:chichiff}) leads to the approximate constraint
\begin{equation}
| g_Z^\chi| \;\lesssim\;
0.56 \left(\frac{1~\mathrm{GeV}}{m_\chi}\right)^{1/2}.
\end{equation}
We emphasize that this bound is considerably weaker than those derived from direct detection and collider searches, and therefore does not impose additional restrictions on the parameter space of interest.

\begin{figure}[htbp!]
\centering
\includegraphics[width=0.65\textwidth]{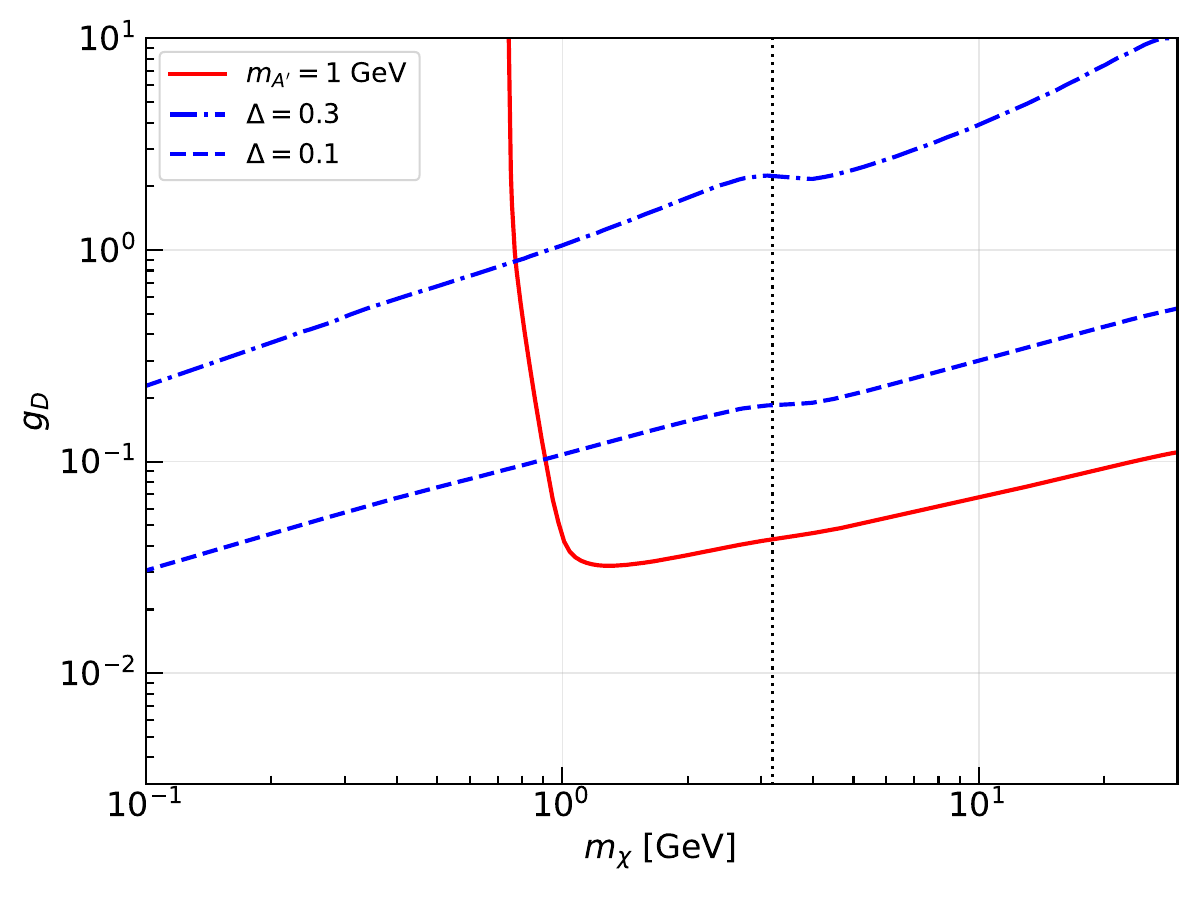}
\caption{\label{fig:DMrelic} The projected contours of the correct DM relic density in the $(m_\chi,g_D)$ plane are shown for $m_{A'} = 1\text{ GeV}$ (solid red), $\Delta = 0.1$ (dashed blue), and $\Delta = 0.3$ (dash-dotted blue). Here, we fix $g_Z^\chi = 0.01$ and $\epsilon = 10^{-6}$. The vertical dotted line denotes the upper bound on the DM mass from direct-detection experiments, assuming $\chi$ constitutes the total observed DM abundance.}
\end{figure}

\section{DM relic density}
\label{sec:dmrelic}
We assume the dark fermion $\chi$ was in thermal equilibrium in the early Universe and froze out to the present-day relic abundance.  
The DM annihilation channels in this model are
(i) $\chi\bar{\chi} \to A'A'$ via $t$- and $u$-channel $\chi$ exchange,
(ii) $\chi\bar{\chi} \to \bar f f$ via $s$-channel $Z$ and $A'$ exchange,  
and (iii) $\chi\bar{\chi} \to \bar t t$ through $t$-channel $\Phi$ exchange.  
Since our analysis focuses on $m_\chi < m_Z/2$, the $\bar t t$ channel is kinematically suppressed and can be neglected.

The thermally averaged annihilation cross section for the channel (i) is
\be
\label{eq:chichiApAp}
\langle \sigma v\rangle_{\chi\bar{\chi} \to A' A'} 
\simeq \frac{\pi \alpha_D^2}{m_\chi^2} 
\frac{\left(1- m_{A'}^2/m_\chi^2\right)^{3/2}}
     {\left(1- 1/2~ m_{A'}^2/m_\chi^2\right)^2},
\ee
where $\alpha_D = g_D^2/(4\pi)$ with $g_D$ being the $U(1)_D$ gauge coupling in the dark sector.
Because this process scales as $g_D^4$, it rapidly becomes dominant and drives the relic abundance of DM below the observed value unless the coupling is small or the channel is kinematically suppressed.

For channel (ii), in the regime $\epsilon \ll g_Z^\chi$, which enhances the LLP signal at the LHC considered in this analysis, the thermally averaged annihilation cross section of $\chi\bar{\chi}$ into SM fermions via the  $Z$ boson exchange dominates and is given in Eq.~\eqref{eq:chichiff}.  
For $m_{A'} \gg m_\chi$, the $A'A'$ channel is closed, and Eq.~\eqref{eq:chichiff} determines the relic abundance.  
However, this process is suppressed by $\left|g_Z^\chi\right|^2$, except near the $Z$ resonance at $m_\chi \simeq m_Z/2$.
Similarly, if $m_\chi \simeq m_{A'}/2$, the resonant $A'$-exchange process can become significant.

 A particularly relevant regime arises when the dark photon is slightly heavier than the dark fermion,
\begin{equation}
\Delta \equiv \frac{m_{A'} - m_\chi}{m_\chi} \ll 1.
\end{equation}
In this \textit{forbidden DM} scenario, annihilation of $\chi\bar{\chi}$ into $A'A'$ remains kinematically allowed but is Boltzmann suppressed at freeze-out. The thermal averaged cross section is approximately~\cite{DAgnolo:2015ujb}
\be
\label{eq:sigmav}
\langle \sigma v\rangle_{\bar{\chi} \chi \to A'A'} \simeq \frac{g_D^4}{2\pi m_\chi^2}  f_\Delta  e^{-2\Delta x},
\ee
where $x \equiv m_\chi/T$ and $f_{\Delta} = [\Delta^{3/2} (2+ \Delta)^{3/2} (2 + \Delta(2+\Delta))]/(1+\Delta)^4$.  
The exponential suppression in Eq.~(\ref{eq:sigmav}) allows sizable $g_D$--required later to enhance dark radiation emission--while still producing the correct relic abundance.

We compute the relic abundance using the \texttt{MicrOMEGAs} package~\cite{Alguero:2023zol} and compare the results with the measured value reported by the Planck collaboration~\cite{Planck:2018vyg}.
Fig.~\ref{fig:DMrelic} shows the resulting contours in the $(m_\chi,\, g_D)$ plane for several benchmark scenarios: the case $m_{A'} = 1\,\text{GeV}$ and the \textit{forbidden DM} regime with mass splittings $\Delta = 0.1$ and $\Delta = 0.3$. Throughout the calculations, we fix $g_Z^\chi = 0.01$, which implies that $m_\chi \gtrsim 3~\text{GeV}$ is excluded by current direct-detection limits, see Fig.~\ref{fig:constraint_mchi_gZchi}, if $\chi$ accounts for the entire observed DM abundance. We also fix $\epsilon = 10^{-6}$ so that the contribution from the s-channel $A'$ mediated is subdominant. 
As seen in Fig.~\ref{fig:DMrelic}, in the \textit{forbidden DM} region, the process $\bar{\chi}\chi \to A' A'$ is exponentially suppressed. Achieving the correct relic abundance, therefore, requires a larger gauge coupling $g_D$ to compensate for the Boltzmann suppression.

\section{Dark Radiation at LLP Detectors}
\label{sec:llp}
In this section, we investigate LLDP signals from dark radiation at several LLP detectors:
FASER/FASER2~\cite{Feng:2017uoz,FASER:2018eoc}, FACET~\cite{Cerci:2021nlb}, and MATHUSLA~\cite{MATHUSLA:2025zyt}. 
The detectors 
FASER/FASER2 and FACET are forward detectors located along the beam collision axis at distances of approximately $480$ m from the ATLAS interaction point (FASER/FASER2) and $119$ m from the CMS interaction point (FACET).
In contrast, MATHUSLA is a large-volume surface detector located transversely above the ATLAS interaction point.
We note that the FASER detector has already been installed and is currently taking data. 

We generate $10^5$ events of $pp \to Z$ at LO using \texttt{MADGRAPH5}~\cite{Alwall:2014hca}, and use \texttt{MadSpin} to decay the on-shell $Z$ boson into a DM pair, $Z \to \bar{\chi}\chi$. Dark radiation of the DM is then simulated with \texttt{PYTHIA 8}~\cite{Sjostrand:2014zea}, which provides the momentum information of the emitted dark photons.
We assume $m_{A'} < 2 m_\chi$, therefore dark photons only decay into SM particles.  
The probability of detecting a dark photon is then computed as 
\be
P_{A'} = 
f(\theta, \phi) 
\int_{L_{\rm min}}^{L_{\rm max}} d \ell 
\frac{e^{-\ell/\ell_{A'}}} {\ell_{A'}} \, \omega \, , 
\label{eq:prob-detection}
\ee
where $L_{\rm min}$ and $L_{\rm max}$ denote the minimal and maximal distances from the interaction point to the detector decay volume along the $(\theta,\phi)$ trajectory, where $\theta$ and $\phi$ are the polar and azimuthal angles of 
the dark photon, respectively. 
The dark photon decay length is $\ell_{A'} = \tau_{A'} |\vec{p}_{A'}|/m_{A'}$, with $\tau_{A'}$ its proper lifetime. The function $f(\theta,\phi)$ encodes the angular acceptance of the detector, and $\omega$ represents additional detector cuts for the final state particles. In our analysis, we simply take $\omega = 1$.

For a cylindrical detector (e.g.\ FASER/FASER2 or FACET) aligned with the beamline and located a distance (d) from the interaction point, the parameters in Eq.~\eqref{eq:prob-detection} are given by
\bea
L_{\rm min} &=& d, \quad 
L_{\rm max} = d+L, \\
f(\theta, \phi) &=& \Theta(R/{L_{\rm min}} - \tan \theta) \,\Theta(\tan \theta - r/L_{\rm max}),
\label{eq:fthetaphi}
\eea
where 
$L$ is the length of the decay volume of the detector, 
$r$ ($R$) is the inner (outer) radius of the decay volume, 
and $\Theta$ is the Heaviside step function. 
For the FACET detector,  
$r={18}$ cm and $R=50$ cm~\cite{Cerci:2021nlb} while 
for the FASER (FASER 2) detector, $r=0$ and $R=10$ (100) cm~\cite{Feng:2017uoz,FASER:2018eoc}. 
Forward LLP detectors are often characterized by pseudorapidity, with acceptance 
$f(\theta, \phi) = \Theta(\eta_{\rm{max}} - \eta_{A'}) \Theta(\eta_{A'} - \eta_{\rm{min}})$. 
For FACET, $\eta_{\rm{min}} \simeq 6$ and 
$\eta_{\rm{max}}\simeq {7.2}$ and for FASER (FASER 2), $\eta_{\rm{min}} \simeq 9$ (7) and $\eta_{\rm{max}} = +\infty$.

For  a box-shaped detector (e.g.\ MATHUSLA) of height (H), width (W), length (L), located at a height (h) above the beamline and horizontal distance (d) from the interaction point, the geometry-dependent quantities are given by~\cite{Du:2021cmt}
\bea
\label{eq:box1-new}
 L_{\rm max} &=& \left\{
\begin{aligned}
&\frac{h+H}{\sin \theta \cos \phi}\quad& {\rm if} \;\;
& \tan \theta > \frac{h + H } { (d+L) \cos \phi}\; \&\; |\tan\phi | < \frac{W}{2(h+H)},\\
&\frac{d+L}{\cos \theta }\quad&  {\rm if} \;\; 
&  \tan \theta < \frac{h + H } { (d+L) \cos \phi}\; \&\; |\sin\phi | < \frac{W}{2(d+L)\tan\theta}, \\
&\frac{W}{2\sin \theta |\sin\phi|}\quad&  {\rm if} \;\; 
& |\sin\phi | > \frac{W}{2(d+L)\tan\theta},
\end{aligned}
\right. \nonumber\\
\eea
\bea
 \label{eq:box2-new}
L_{\rm min} &=& \left\{
\begin{aligned}
&\frac{ h }{\sin \theta \cos \phi}\quad& {\rm if} \;\; & \tan \theta < \frac{h } { d \cos \phi}, \\
&\frac{d}{\cos \theta }\quad&  {\rm if} \;\; &  \tan \theta > \frac{h} {d\cos \phi},
\end{aligned}
\right.  
\eea
\bea
\label{eq:box3-new}
f(\theta, \phi) &=&  \Theta \left( \tan \theta - \frac{h}{(d+L)\cos \phi} \right) \, 
 \Theta \left(  \frac{h + H }{d\cos \phi} -  \tan \theta\right) \, 
 \Theta \left(\frac{W}{2 h} - |\tan \phi| \right)  \Theta \left(\cos\phi \right). \nonumber  \\
\eea
For the MATHUSLA detector, we take the latest geometry proposal 
$d=$ 70 m, 
$h=$ 81 m, 
$W=$ 40 m, 
$L=$ 40 m, 
and $H=$ 11 m 
\cite{MATHUSLA:2025zyt}.

The expected number of detectable dark photon events in far detectors is given by
\be
N = {\cal L} \cdot \sigma_{A'} \cdot \langle P_{A'} \rangle  \quad {\rm with }  \quad
\langle P_{A'} \rangle = \frac{1}{N_{\rm A'}} \sum_{i=1}^{N_{A'}} P_{A'_i}, 
\ee
where ${\cal L}$ is the integrated luminosity, $\sigma_{A'} = \sigma_{pp \to Z} \times \text{BR}(Z\to\bar{\chi} \chi) \times \langle {n}_{A'}\rangle$ is the total dark photon production cross section,  $\langle{n}_{A'}\rangle$ denotes the expectation value for the number of dark photons radiated from a $\bar{\chi} \chi$ pair, 
$\langle P_{A'} \rangle$ is the average detection probability,
$N_{A'}$ is the number of dark photons produced in the simulation,
and $P_{A'_i}$ is the detection probability of the $i$-th event, computed using Eq.~\eqref{eq:prob-detection}.

\begin{figure}[htbp!]
\centering
\includegraphics[width=0.65\textwidth]{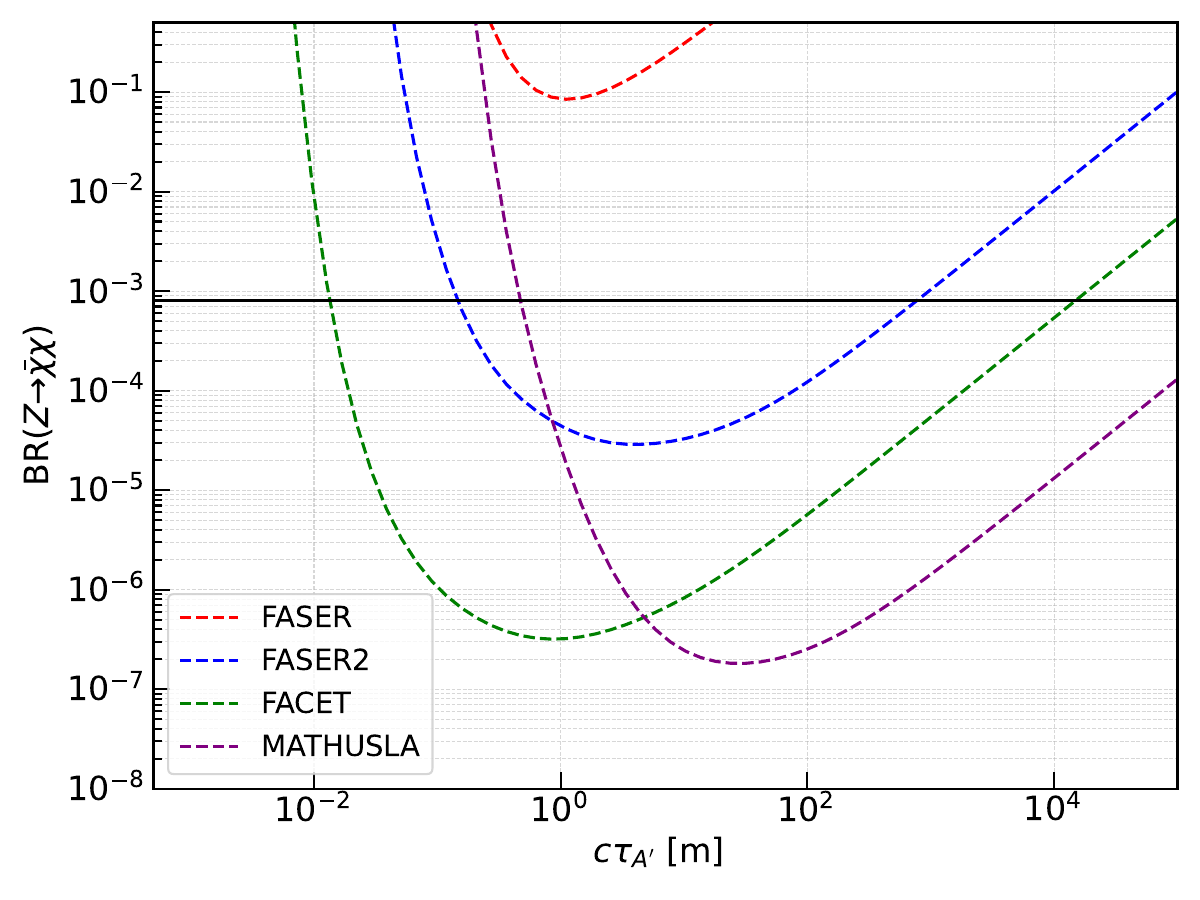}
\caption{\label{fig:reach_BRZchichi}
Projected $95\%$ exclusion sensitivity for FASER (dashed red), FASER2 (dashed blue), FACET (dashed green), and MATHUSLA (dashed purple) in the plane of ${\rm BR}(Z\to\bar{\chi}\chi)$ and the proper lifetime of the dark photon.  
We fix $m_\chi = 2$~GeV and $\Delta = 0.3$.  
The horizontal solid black line shows the current constraint from the $Z$ invisible decay width.
} 
\end{figure}

\begin{figure}[htbp!]
\centering
\includegraphics[width=0.65\textwidth]{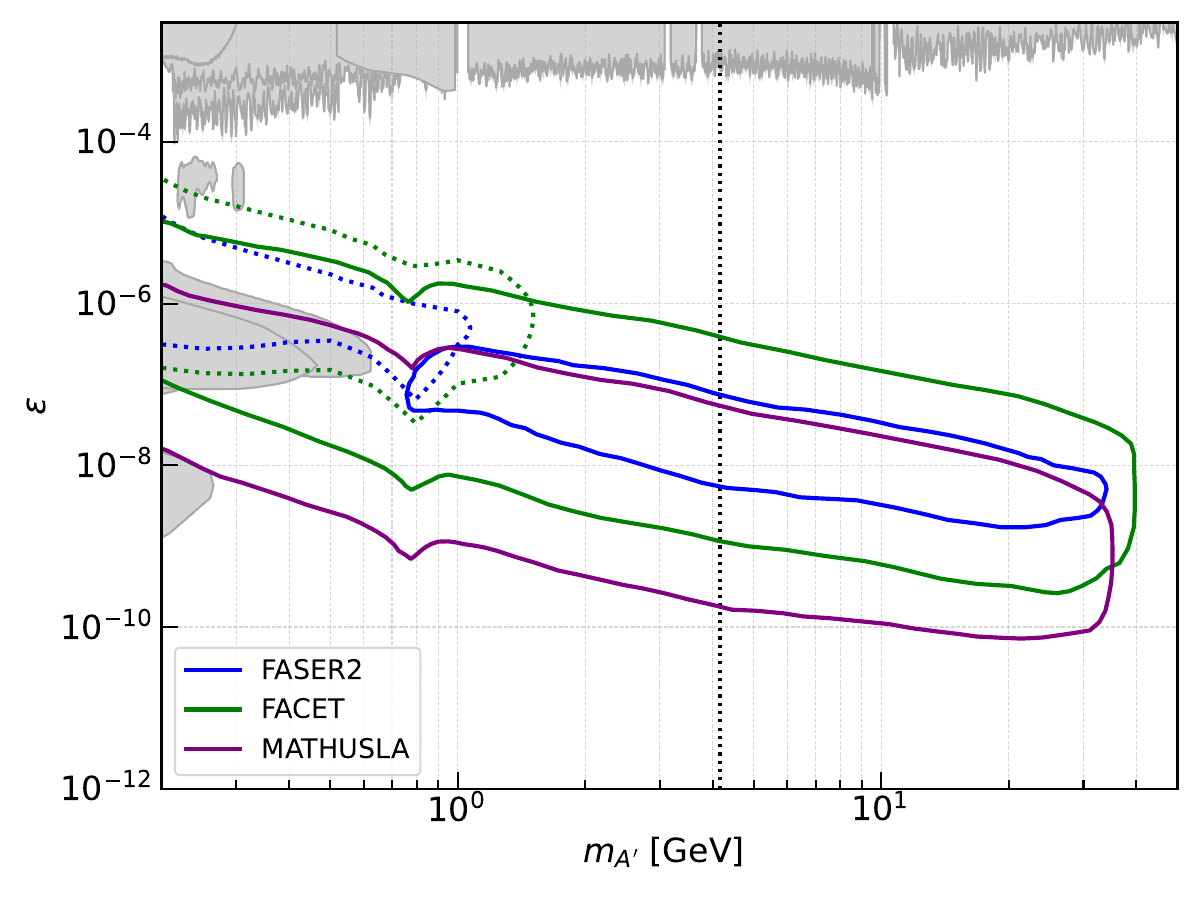}
\caption{\label{fig:reach_mAp_eps}
Projected $95\%$ exclusion sensitivity for FASER2, FACET, and MATHUSLA in the $(m_{A'},\epsilon)$ plane.  
We fix $\Delta=0.3$, assume ${\rm BR}(Z\to\bar{\chi}\chi)=10^{-4}$, and vary $g_D$ to reproduce the observed DM relic abundance.  
Solid green, blue, and purple contours correspond to the dark-radiation induced dark photon signals at FASER2, FACET, and MATHUSLA, respectively.  
The dotted green and blue curves show the sensitivities to the conventional dark photon signals from meson decay and proton bremsstrahlung, computed using the {\tt FORESEE} package~\cite{Kling:2021fwx}.  
The vertical dotted line indicates the current DM direct detection bound on $m_\chi$ discussed in Sec.~\ref{sec:exp_constraint}.  
Light gray regions denote existing constraints on the conventional dark photon parameter space.
}
\end{figure}

\section{Numerical results}
\label{sec:result}

In this section, we present the projected sensitivities of FASER2, FACET, and MATHUSLA to LLDPs.  
We assume zero background and set the $95\%$ exclusion reach by requiring $N=3$ signal events.  
For FASER2, FACET, and MATHUSLA we take an integrated luminosity of ${\cal L}=3~\text{ab}^{-1}$, while the LHC run 2 luminosity for FASER is ${\cal L}=0.15~\text{ab}^{-1}$. 

Fig.~\ref{fig:reach_BRZchichi} displays the projected sensitivity in the $( {\rm BR}(Z\to\bar{\chi}\chi), c\tau_{A'} )$ plane for a benchmark point with $m_\chi = 2~\text{GeV}$ and $\Delta = 0.3$, which requires $g_D \simeq 1.7$ to obtain the correct relic abundance.  
While FASER alone does not surpass the current $Z$ invisible width bound, FASER2, FACET, and MATHUSLA improve significantly upon it.  
FASER2 is sensitive down to ${\rm BR}(Z\to\bar{\chi}\chi)\simeq 3\times10^{-5}$ at $c\tau_{A'}\sim 4$~m, while FACET reaches two orders of magnitude lower, ${\rm BR}\simeq 3\times10^{-7}$ near $c\tau_{A'}\sim 1$~m.  
Thanks to its large fiducial volume and transverse geometry, MATHUSLA provides the strongest reach, probing ${\rm BR}\simeq 2\times10^{-7}$ at $c\tau_{A'}\sim30$~m.  
Although FASER and FACET are located further from the interaction point, dark photons produced in the forward region typically carry larger momenta and hence larger boost factors, requiring shorter proper lifetimes to decay inside the detector as compared to the transverse region detector, {\it e.g.} MATHUSLA.

Fig.~\ref{fig:reach_mAp_eps} presents the sensitivities in the $(m_{A'},\epsilon)$ plane for ${\rm BR}(Z\to\bar{\chi}\chi)=10^{-4}$ and $\Delta=0.3$.  
We find that dark photons originating from dark radiation (solid contours) provide significantly enhanced reach compared to the conventional production mechanisms (dotted contours).  
For FASER2, the dark radiation signal allows probing $m_{A'}\sim(0.8$--$30$~GeV) and $\epsilon\simeq$($10^{-9}$--$3\times10^{-7}$), whereas the meson/bremsstrahlung channels are restricted to $m_{A'}\lesssim1$~GeV and $\epsilon\gtrsim7\times10^{-8}$.  
Similarly, FACET can test masses up to $\sim40$~GeV and mixings as small as $\epsilon\sim 2\times10^{-10}$ for dark radiation, while the conventional channels are limited to $m_{A'}\lesssim1.5$~GeV and $\epsilon\gtrsim3\times10^{-8}$.  
Notably, for MATHUSLA the entire parameter space from conventional production is already excluded by existing beam-dump experiments, but dark radiation opens a large new region at small $\epsilon$ and heavy $m_{A'}$.

At small $m_{A'}$, relic-density requirements force $g_D$ to be small, reducing the dark-radiation yield.  
The dips near $m_{A'}\sim0.8$~GeV arise from hadronic resonances in the dark photon width.  
The vertical dotted line reflects direct-detection limits, which eliminate part of the parameter space unless $\chi$ constitutes only a subcomponent of the DM density.

\section{Conclusion}
\label{sec:conclusion}

In this work, we present a framework in which dark photons can be produced via the radiation from dark matter $\chi$ in $Z$ decay, providing an efficient probe of LLDP with long-lived particle detectors at the LHC. In contrast to traditional searches for LLDPs from meson decays and bremsstrahlung, parameters governing the production and decay of dark photons are decoupled in this scenario. The interaction between dark matter and the 
$Z$ boson can be induced at one-loop level via a heavy colored scalar 
$\Phi$ that couples to both dark matter and Standard Model fermions, chosen to be the top quark in this study.
A sizable effective coupling $g_Z^\chi$ can be generated through $(t,\Phi)$ loops and allows the decay $Z\to\chi\chi$ to occur at rates close to current experimental limits.
The dark matter subsequently emits dark radiation, $\chi\to\chi A'$, producing an energetic source of LLDPs whose production is governed by the $U(1)_D$ gauge coupling that can be determined by thermal relic considerations of the dark sector.

We analyzed the thermal history of $\chi$, highlighting the importance of the forbidden-annihilation regime, which naturally accommodates large dark gauge couplings while maintaining the correct relic abundance.
We then evaluated leading constraints from the invisible $Z$ width, monojets, and isospin-violating DM direct detection, identifying the viable region in the $(m_\chi,g_Z^\chi,g_D)$ parameter space.
Using detailed simulations that incorporate realistic detector geometry and $A'$ decay kinematics, we study the current constraints and projected sensitivities of FASER/FASER2, FACET, and MATHUSLA.
We found that production of dark photons from dark matter radiation dramatically enhances LLP sensitivity compared to the traditional dark photon production channels.
The future FASER2 can test branching ratios down to ${\rm BR}(Z\to\bar{\chi}\chi)\sim10^{-5}$ and probe dark photon masses up to $30$ GeV with kinetic mixings down to $\epsilon\sim10^{-7}$.
FACET and MATHUSLA achieve the strongest coverages, extending the sensitivity to ${\rm BR}(Z\to\bar{\chi}\chi)\sim{\cal O}(10^{-7})$ and exploring vast regions of parameter space that are inaccessible in the conventional dark photon models.

\section*{Acknowledgments}
We would like to thank Tzu-Chiang Yuan and Shu-Yu Ho for useful discussions. This work was partially supported by the National Science and Technology Council (NSTC) of Taiwan under Grant No. NSTC-113-2112-M-003-007 and NSTC-114-2112-M-003-009 (CRC), the Ministry of Education (Higher Education Sprout Project NTU-114L104022-1), and the National Center for Theoretical Sciences of Taiwan (VQT).
This work was also supported in part by the Vietnam National Foundation for Science and Technology Development (NAFOSTED) under grant number 103.01-2023.50 (VQT).

\appendix
\label{appendix}

\section{Constraints from Higgs Signal Strength Measurements}
\label{app:constraints_colored_scalar}
For the case $m_h < 2 m_\Phi$ and moderate values of $\lambda_{H\Phi}$, the modification to the Higgs total width is negligible. The leading-order correction to Higgs production via gluon fusion relative to the SM can be expressed as~\cite{Batell:2011pz}
\be
\label{eq:ggh}
\frac{\sigma(gg \to h)}{\sigma(gg \to h)_{\rm SM}} \simeq
\frac{\Gamma(h \to gg)}{\Gamma(h \to gg)_{\rm SM}}
= \Bigg| 1 - \frac{1}{2}\frac{\lambda_{H\Phi} v^2}{m_{\Phi}^2} C_\Phi 
\frac{A_0(\tau_\Phi)}{\sum_f A_{1/2} (\tau_f)} \Bigg|^2 ,
\ee
where $\tau_i = m_h^2/(4 m_i^2)$, $C_\Phi = 4/3$ is the quadratic Casimir of the $SU(3)_C$ fundamental representation, and $A_0$ ($A_{1/2}$) denotes the scalar (fermion) loop functions (see below).

Similarly, the Higgs partial width to photons is modified as
\be
\label{eq:hgaga}
\frac{\Gamma(h \to \gamma\gamma)}{\Gamma(h \to \gamma\gamma)_{\rm SM}}
= \Bigg| 1 + \frac{1}{2}\frac{\lambda_{H\Phi} v^2}{2m_{\Phi}^2} d_\Phi Q_\Phi^2 
\frac{A_0(\tau_\Phi)}{A_1(\tau_W) - \sum_f N_f Q_f^2 A_{1/2}(\tau_f)} \Bigg|^2 ,
\ee
where $d_\Phi = 3$ is the dimension of the $SU(3)_C$ representation, $Q_\Phi$ is the electric charge of $\Phi$, $N_f$ and $Q_f$ denote the color factor and charge of fermion $f$. .

The loop functions for $h \to gg$ and $h \to \gamma\gamma$ are given as 
\bea
A_0(\tau) = \frac{1}{\tau^2} \left(f(\tau) -\tau \right), \\
A_{1/2}(\tau) = \frac{2}{\tau^2} \left[ \tau + (\tau -1) f(\tau) \right], \\
A_1(\tau) = 2 + \frac{3}{\tau} + \frac{3}{\tau} (2 - \frac{1}{\tau}) f(\tau), 
\eea
where $\tau = m_h^2/(4 m^2)$, with $m$ being the mass of particle in the loop, and the function $f(\tau)$ is given as 
\be
\label{ftau}
f( \tau ) =  \left\{ 
\begin{array}{cr}
  {\rm arcsin}^2 (\sqrt{\tau})  \, , &   (\tau \leq 1) \, , \\
- \frac{1}{4} \left[ \log \left( \frac{1 + \sqrt{1 - \tau^{-1}}} {1 - \sqrt{1 - \tau^{-1}}}\right) - i \pi \right]^2 \, ,  & (\tau >1) \, ; \end{array} \right.
\ee

To quantify the impact on Higgs searches, we compute the signal strength for the diphoton channel,
\be
\mu^{\gamma\gamma}_{ggh} = \frac{\sigma(gg \to h) \times {\rm BR}(h\to \gamma \gamma)}{\sigma(gg \to h)_{\rm SM} \times {\rm BR}(h\to \gamma \gamma)_{\rm SM}}
\simeq \frac{\Gamma(h \to gg)}{\Gamma(h \to gg)_{\rm SM}} \times \frac{\Gamma(h \to \gamma\gamma)}{\Gamma(h \to \gamma\gamma)_{\rm SM}}.
\ee
At leading order, the Higgs decay widths into fermion pairs, $WW^*$, and $ZZ^*$ remain unchanged compared to the SM. Thus, the corresponding signal strengths for these modes, with the Higgs produced via gluon fusion, reduce to
\be
\mu^{f\bar{f}, WW^*, ZZ^*}_{ggh} \simeq \frac{\Gamma(h \to gg)}{\Gamma(h \to gg)_{\rm SM}}.
\ee

\begin{figure}[htbp!]
\centering
\includegraphics[width=0.65\textwidth]{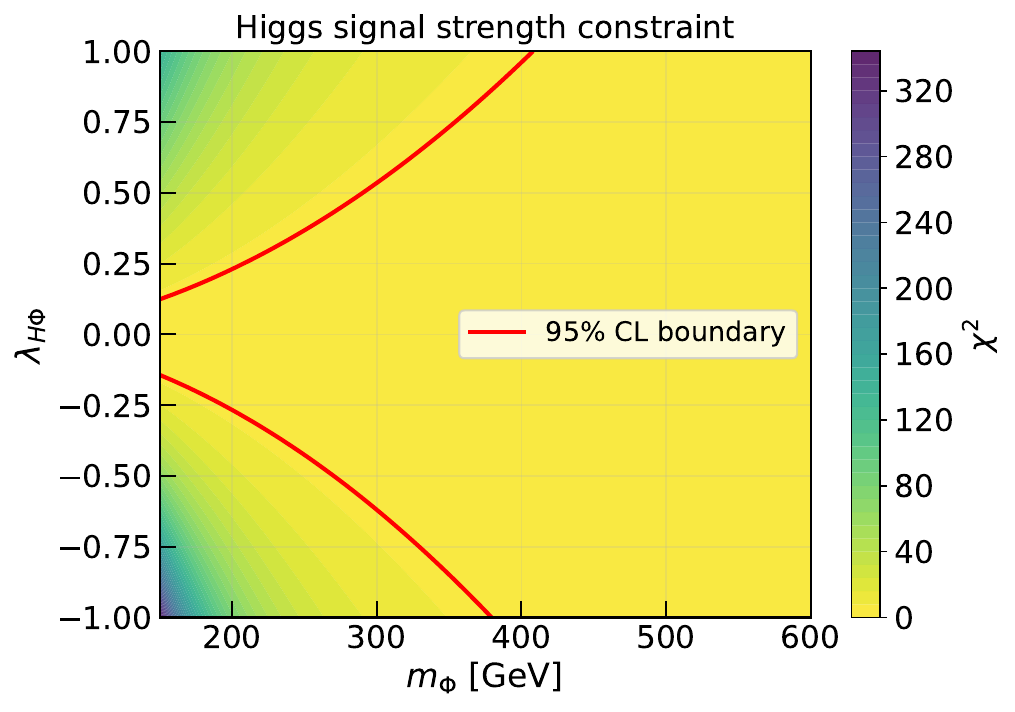}
\caption{\label{fig:higgsconstraint} The constraints from the Higgs signal strength data projected on $m_{\Phi}, \lambda_{H\Phi}$ plane. }
\end{figure}

Using the latest ATLAS measurements of Higgs signal strengths of Higgs decays to $\gamma\gamma$, $WW^*$, $ZZ^*$, $\tau^+\tau^-$ and $\mu^+\mu^-$ modes from Ref.~\cite{ATLAS:2022vkf}, we derive the constraints on the colored scalar mass and the portal coupling $\lambda_{H\Phi}$, as shown in Fig.~\ref{fig:higgsconstraint}. For $|\lambda_{H\Phi}| \gtrsim 1$, the colored scalar mass must be heavier than about 400 GeV. This bound is weaker than the constraints obtained from searches for top quark pairs plus missing energy at the LHC, as discussed previously.

On the other hand, if the colored scalar is light, the quadratic interaction $hh\Phi\Phi^*$ can enhance the di-Higgs production at $pp$ collision~\cite{Kribs:2012kz}. Given a heavy $\Phi$ above the TeV scale in this analysis, these contributions to Higgs properties are not significant and satisfied the current experimental data.

\section{Form Factors For One-Loop Effective \texorpdfstring{$Z\bar{\chi}\chi$}{Z chi chi}}
\label{app:formfactors}
The expressions of the form factors in Eq.~(\ref{eq:Zeff}) are given by 
\begin{align}
\label{eq:formfactors}
F_L &= \frac{-i}{16 \pi^2} N_c |\lambda_{\chi t}|^2  \Bigg\{ g_R^{t} \left[ - B_0 + 2 C_{00} - m_t^2  C_0  
+ m_\chi^2 C_{11} + q^2 \left( C_0  + C_1 + C_2\right) \right] \nonumber \\
& \quad + g_L^{t} m_t^2 C_0   + g_{Z\Phi} \left[ 2 \widetilde{C}_{00}  
+ m_\chi^2 \left(  \widetilde{C}_{1}  +  \widetilde{C}_{11}  \right)  \right] \Bigg\}, 
\end{align}
\begin{align}
F_R &=  \frac{-i}{16 \pi^2} N_c |\lambda_{\chi t}|^2 m_\chi^2 \left[ g_R^{t}  C_{11} + g_{Z\Phi} \left(  \widetilde{C}_{1}  +  \widetilde{C}_{11}  \right)\right], \\
F_M &=  \frac{i}{16 \pi^2} N_c |\lambda_{\chi t}|^2 m_\chi^2  \left[g_R^{t}  \left(C_{11} + C_{1} \right)  + g_{Z\Phi}\left(  \widetilde{C}_{1}  +  \widetilde{C}_{11}  \right)\right],  \\
F_E &= 0. 
\end{align}
where $N_c$ is color factor,  
\begin{align}
              g_L^t &= \frac{g}{c_W} \left(\frac{1}{2} - \frac{2}{3}s_W^2\, \right),  
              \quad\quad g_R^t =  \frac{g}{c_W} \left(-\frac{2}{3}s_W^2\, \right),
              \label{eq:gLt}
\end{align}
are left-handed and right-handed components of $Z\bar{t}t$ vertex, respectively and 
$g_{Z\Phi} = \frac{g}{c_W}\left(-\frac{2}{3} s_W^2\right)$ is the coupling of $Z\Phi\Phi^*$ vertex.
In Eq.~(\ref{eq:formfactors}), $B_0$, $C_{j,jk}$ and $\widetilde{C}_{j,jk}$ are Passarino-Veltman (PV) coefficient functions~\cite{Passarino:1978jh} which are given as 
\begin{align}
B_0 &\equiv \mathbf{B}_0(m_\chi^2;\,m_t^2,\,m_{\Phi}^2), \\
C_{j,jk} &\equiv \mathbf{C}_{j,jk}(q^2,m_\chi^2,m_\chi^2; m_t^2,m_t^2,m_{\Phi}^2),  \\
\widetilde{C}_{j,jk} &\equiv \mathbf{C}_{j,jk}(q^2,m_\chi^2,m_\chi^2; m_{\Phi}^2, m_{\Phi}^2,m_t^2). 
\end{align}

We now examine the UV divergence structure of the induced coupling.
Working in dimensional regularization with $d = 4 - 2\epsilon$
spacetime dimensions, the relevant PV scalar functions carry the following divergent parts are
\begin{align}
    {\rm Div}[B_0] = \Delta_\epsilon\,,
    \qquad
    {\rm Div}[C_{00}]
    = {\rm Div}[\widetilde{C}_{00}]
    = \frac{ \Delta_\epsilon}{4}\,,    \label{eq:divparts}
\end{align}
where $ \Delta_\epsilon \equiv \frac{1}{\epsilon} - \gamma_E + \ln(4\pi)$ is the
standard UV regulator. Thus, only the $F_L$ form factor contains the divergent parts with
\begin{align}
    {\rm Div}[F_L] &\sim \Bigg[ g_R^{t} \left( - {\rm Div}[B_0] + 2 \, {\rm Div}[C_{00}]  \right) +   g_{Z\Phi} \left(2 \, {\rm Div}[\widetilde{C}_{00}] \right)  \Bigg],  \\
    &\sim  \frac{\Delta_\epsilon}{2}  \left( g_{Z\Phi} - g_R^{t} \right)  = 0. 
\end{align}
Therefore, the UV divergence cancels exactly.

\begin{figure}[t]
\centering
\includegraphics[width=0.49\textwidth]{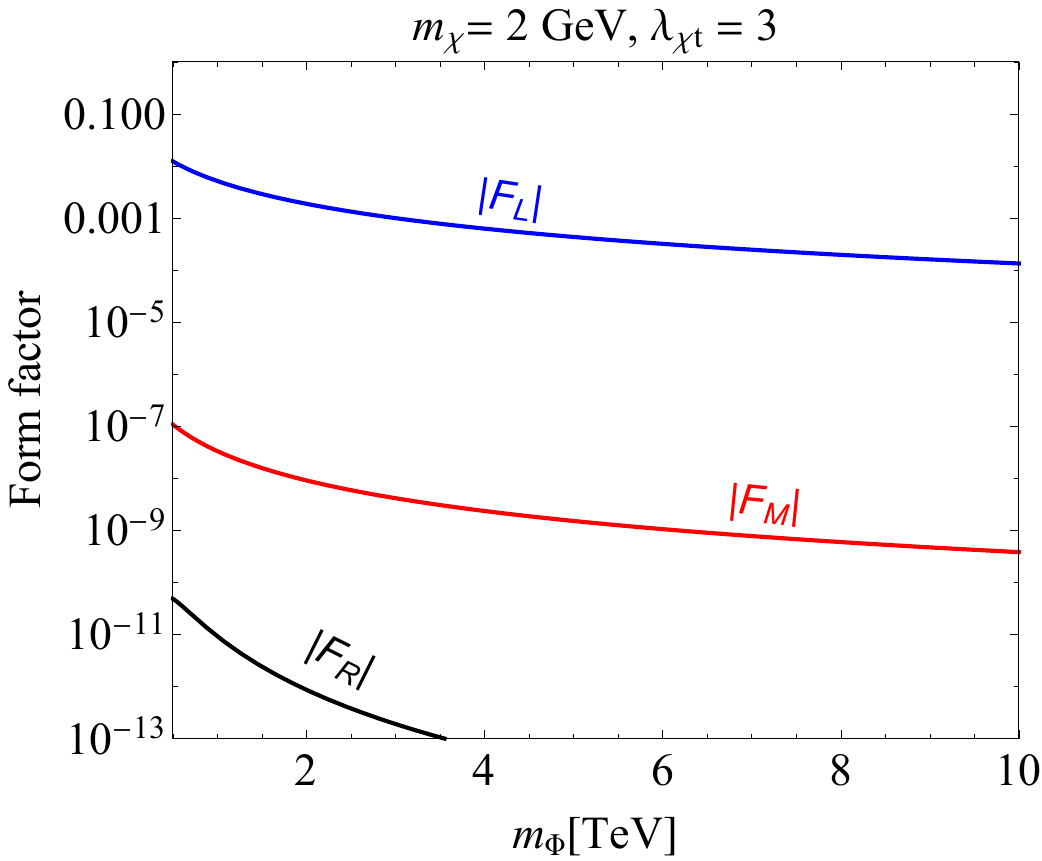}
\includegraphics[width=0.47\textwidth]{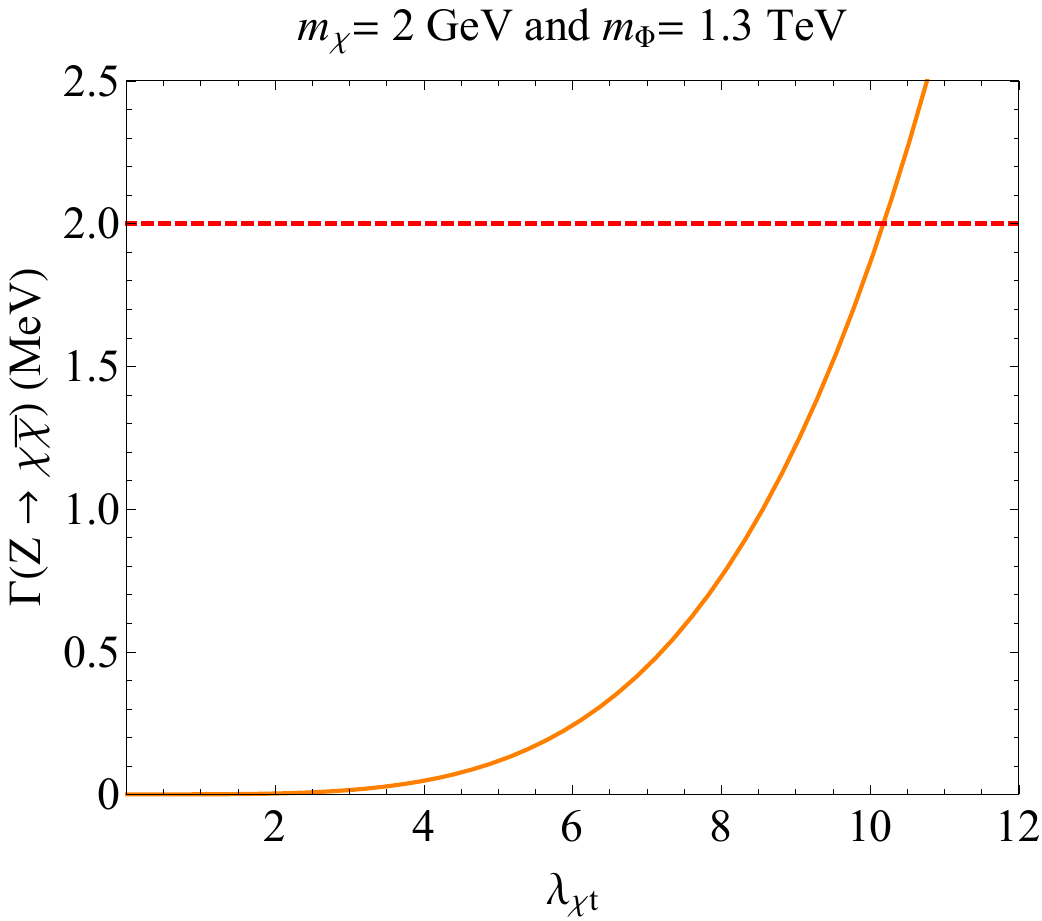}
\caption{Left panel: Form factors as a function of colored scalar mass with fixing $m_\chi = 2$ GeV and $\lambda_{\chi t} = 3$. The solid blue, black and red lines represent $|F_L|$, $|F_R|$ and $|F_M|$, respectively. Right panel: The decay width of $Z\chi \bar{\chi}$  (orange line) as a function of the Yukawa coupling $\lambda_{\chi t}$ with fixing $m_\chi = 2$ GeV and $m_\Phi = 1.3$ TeV. The dashed red line indicates the current bound from the invisible decay of the $Z$ boson.}
\label{fig:Zloop}
\end{figure}

As shown in the left panel of Fig.~\ref{fig:Zloop}, the form factor $F_L$ dominates over others in our parameter space of interest. With fixing $m_\chi = 2$ GeV and $m_\Phi = 1.3$ TeV, the current constraint from the $Z$ invisible constraint requires the Yukawa coupling to be $\lambda_{\chi t} \lesssim 10$ as shown in the right panel of Fig.~\ref{fig:Zloop}.

\allowdisplaybreaks
\bibliographystyle{apsrev4-1}
\bibliography{refs}

\begin{thebibliography}{47}%
\makeatletter
\providecommand \@ifxundefined [1]{%
 \@ifx{#1\undefined}
}%
\providecommand \@ifnum [1]{%
 \ifnum #1\expandafter \@firstoftwo
 \else \expandafter \@secondoftwo
 \fi
}%
\providecommand \@ifx [1]{%
 \ifx #1\expandafter \@firstoftwo
 \else \expandafter \@secondoftwo
 \fi
}%
\providecommand \natexlab [1]{#1}%
\providecommand \enquote  [1]{``#1''}%
\providecommand \bibnamefont  [1]{#1}%
\providecommand \bibfnamefont [1]{#1}%
\providecommand \citenamefont [1]{#1}%
\providecommand \href@noop [0]{\@secondoftwo}%
\providecommand \href [0]{\begingroup \@sanitize@url \@href}%
\providecommand \@href[1]{\@@startlink{#1}\@@href}%
\providecommand \@@href[1]{\endgroup#1\@@endlink}%
\providecommand \@sanitize@url [0]{\catcode `\\12\catcode `\$12\catcode
  `\&12\catcode `\#12\catcode `\^12\catcode `\_12\catcode `\%12\relax}%
\providecommand \@@startlink[1]{}%
\providecommand \@@endlink[0]{}%
\providecommand \url  [0]{\begingroup\@sanitize@url \@url }%
\providecommand \@url [1]{\endgroup\@href {#1}{\urlprefix }}%
\providecommand \urlprefix  [0]{URL }%
\providecommand \Eprint [0]{\href }%
\providecommand \doibase [0]{http://dx.doi.org/}%
\providecommand \selectlanguage [0]{\@gobble}%
\providecommand \bibinfo  [0]{\@secondoftwo}%
\providecommand \bibfield  [0]{\@secondoftwo}%
\providecommand \translation [1]{[#1]}%
\providecommand \BibitemOpen [0]{}%
\providecommand \bibitemStop [0]{}%
\providecommand \bibitemNoStop [0]{.\EOS\space}%
\providecommand \EOS [0]{\spacefactor3000\relax}%
\providecommand \BibitemShut  [1]{\csname bibitem#1\endcsname}%
\let\auto@bib@innerbib\@empty
\bibitem [{\citenamefont {Pospelov}\ \emph {et~al.}(2008)\citenamefont
  {Pospelov}, \citenamefont {Ritz},\ and\ \citenamefont
  {Voloshin}}]{Pospelov:2007mp}%
  \BibitemOpen
  \bibfield  {author} {\bibinfo {author} {\bibfnamefont {M.}~\bibnamefont
  {Pospelov}}, \bibinfo {author} {\bibfnamefont {A.}~\bibnamefont {Ritz}}, \
  and\ \bibinfo {author} {\bibfnamefont {M.~B.}\ \bibnamefont {Voloshin}},\
  }\href {\doibase 10.1016/j.physletb.2008.02.052} {\bibfield  {journal}
  {\bibinfo  {journal} {Phys. Lett. B}\ }\textbf {\bibinfo {volume} {662}},\
  \bibinfo {pages} {53} (\bibinfo {year} {2008})},\ \Eprint
  {http://arxiv.org/abs/0711.4866} {arXiv:0711.4866 [hep-ph]} \BibitemShut
  {NoStop}%
\bibitem [{\citenamefont {D'Agnolo}\ and\ \citenamefont
  {Ruderman}(2015)}]{DAgnolo:2015ujb}%
  \BibitemOpen
  \bibfield  {author} {\bibinfo {author} {\bibfnamefont {R.~T.}\ \bibnamefont
  {D'Agnolo}}\ and\ \bibinfo {author} {\bibfnamefont {J.~T.}\ \bibnamefont
  {Ruderman}},\ }\href {\doibase 10.1103/PhysRevLett.115.061301} {\bibfield
  {journal} {\bibinfo  {journal} {Phys. Rev. Lett.}\ }\textbf {\bibinfo
  {volume} {115}},\ \bibinfo {pages} {061301} (\bibinfo {year} {2015})},\
  \Eprint {http://arxiv.org/abs/1505.07107} {arXiv:1505.07107 [hep-ph]}
  \BibitemShut {NoStop}%
\bibitem [{\citenamefont {Aaij}\ \emph
  {et~al.}(2020{\natexlab{a}})\citenamefont {Aaij} \emph
  {et~al.}}]{LHCb:2019vmc}%
  \BibitemOpen
  \bibfield  {author} {\bibinfo {author} {\bibfnamefont {R.}~\bibnamefont
  {Aaij}} \emph {et~al.} (\bibinfo {collaboration} {LHCb}),\ }\href {\doibase
  10.1103/PhysRevLett.124.041801} {\bibfield  {journal} {\bibinfo  {journal}
  {Phys. Rev. Lett.}\ }\textbf {\bibinfo {volume} {124}},\ \bibinfo {pages}
  {041801} (\bibinfo {year} {2020}{\natexlab{a}})},\ \Eprint
  {http://arxiv.org/abs/1910.06926} {arXiv:1910.06926 [hep-ex]} \BibitemShut
  {NoStop}%
\bibitem [{\citenamefont {Aaij}\ \emph
  {et~al.}(2020{\natexlab{b}})\citenamefont {Aaij} \emph
  {et~al.}}]{LHCb:2020ysn}%
  \BibitemOpen
  \bibfield  {author} {\bibinfo {author} {\bibfnamefont {R.}~\bibnamefont
  {Aaij}} \emph {et~al.} (\bibinfo {collaboration} {LHCb}),\ }\href {\doibase
  10.1007/JHEP10(2020)156} {\bibfield  {journal} {\bibinfo  {journal} {JHEP}\
  }\textbf {\bibinfo {volume} {10}},\ \bibinfo {pages} {156} (\bibinfo {year}
  {2020}{\natexlab{b}})},\ \Eprint {http://arxiv.org/abs/2007.03923}
  {arXiv:2007.03923 [hep-ex]} \BibitemShut {NoStop}%
\bibitem [{\citenamefont {Sirunyan}\ \emph {et~al.}(2020)\citenamefont
  {Sirunyan} \emph {et~al.}}]{CMS:2019buh}%
  \BibitemOpen
  \bibfield  {author} {\bibinfo {author} {\bibfnamefont {A.~M.}\ \bibnamefont
  {Sirunyan}} \emph {et~al.} (\bibinfo {collaboration} {CMS}),\ }\href
  {\doibase 10.1103/PhysRevLett.124.131802} {\bibfield  {journal} {\bibinfo
  {journal} {Phys. Rev. Lett.}\ }\textbf {\bibinfo {volume} {124}},\ \bibinfo
  {pages} {131802} (\bibinfo {year} {2020})},\ \Eprint
  {http://arxiv.org/abs/1912.04776} {arXiv:1912.04776 [hep-ex]} \BibitemShut
  {NoStop}%
\bibitem [{\citenamefont {Hayrapetyan}\ \emph {et~al.}(2023)\citenamefont
  {Hayrapetyan} \emph {et~al.}}]{CMS:2023hwl}%
  \BibitemOpen
  \bibfield  {author} {\bibinfo {author} {\bibfnamefont {A.}~\bibnamefont
  {Hayrapetyan}} \emph {et~al.} (\bibinfo {collaboration} {CMS}),\ }\href
  {\doibase 10.1007/JHEP12(2023)070} {\bibfield  {journal} {\bibinfo  {journal}
  {JHEP}\ }\textbf {\bibinfo {volume} {12}},\ \bibinfo {pages} {070} (\bibinfo
  {year} {2023})},\ \Eprint {http://arxiv.org/abs/2309.16003} {arXiv:2309.16003
  [hep-ex]} \BibitemShut {NoStop}%
\bibitem [{\citenamefont {Abreu}\ \emph {et~al.}(2024)\citenamefont {Abreu}
  \emph {et~al.}}]{FASER:2023tle}%
  \BibitemOpen
  \bibfield  {author} {\bibinfo {author} {\bibfnamefont {H.}~\bibnamefont
  {Abreu}} \emph {et~al.} (\bibinfo {collaboration} {FASER}),\ }\href {\doibase
  10.1016/j.physletb.2023.138378} {\bibfield  {journal} {\bibinfo  {journal}
  {Phys. Lett. B}\ }\textbf {\bibinfo {volume} {848}},\ \bibinfo {pages}
  {138378} (\bibinfo {year} {2024})},\ \Eprint
  {http://arxiv.org/abs/2308.05587} {arXiv:2308.05587 [hep-ex]} \BibitemShut
  {NoStop}%
\bibitem [{\citenamefont {Buschmann}\ \emph {et~al.}(2015)\citenamefont
  {Buschmann}, \citenamefont {Kopp}, \citenamefont {Liu},\ and\ \citenamefont
  {Machado}}]{Buschmann:2015awa}%
  \BibitemOpen
  \bibfield  {author} {\bibinfo {author} {\bibfnamefont {M.}~\bibnamefont
  {Buschmann}}, \bibinfo {author} {\bibfnamefont {J.}~\bibnamefont {Kopp}},
  \bibinfo {author} {\bibfnamefont {J.}~\bibnamefont {Liu}}, \ and\ \bibinfo
  {author} {\bibfnamefont {P.~A.~N.}\ \bibnamefont {Machado}},\ }\href
  {\doibase 10.1007/JHEP07(2015)045} {\bibfield  {journal} {\bibinfo  {journal}
  {JHEP}\ }\textbf {\bibinfo {volume} {07}},\ \bibinfo {pages} {045} (\bibinfo
  {year} {2015})},\ \Eprint {http://arxiv.org/abs/1505.07459} {arXiv:1505.07459
  [hep-ph]} \BibitemShut {NoStop}%
\bibitem [{\citenamefont {Du}\ \emph {et~al.}(2022)\citenamefont {Du},
  \citenamefont {Fang}, \citenamefont {Liu},\ and\ \citenamefont
  {Tran}}]{Du:2021cmt}%
  \BibitemOpen
  \bibfield  {author} {\bibinfo {author} {\bibfnamefont {M.}~\bibnamefont
  {Du}}, \bibinfo {author} {\bibfnamefont {R.}~\bibnamefont {Fang}}, \bibinfo
  {author} {\bibfnamefont {Z.}~\bibnamefont {Liu}}, \ and\ \bibinfo {author}
  {\bibfnamefont {V.~Q.}\ \bibnamefont {Tran}},\ }\href {\doibase
  10.1103/PhysRevD.105.055012} {\bibfield  {journal} {\bibinfo  {journal}
  {Phys. Rev. D}\ }\textbf {\bibinfo {volume} {105}},\ \bibinfo {pages}
  {055012} (\bibinfo {year} {2022})},\ \Eprint
  {http://arxiv.org/abs/2111.15503} {arXiv:2111.15503 [hep-ph]} \BibitemShut
  {NoStop}%
\bibitem [{\citenamefont {Du}\ \emph {et~al.}(2020)\citenamefont {Du},
  \citenamefont {Liu},\ and\ \citenamefont {Tran}}]{Du:2019mlc}%
  \BibitemOpen
  \bibfield  {author} {\bibinfo {author} {\bibfnamefont {M.}~\bibnamefont
  {Du}}, \bibinfo {author} {\bibfnamefont {Z.}~\bibnamefont {Liu}}, \ and\
  \bibinfo {author} {\bibfnamefont {V.~Q.}\ \bibnamefont {Tran}},\ }\href
  {\doibase 10.1007/JHEP05(2020)055} {\bibfield  {journal} {\bibinfo  {journal}
  {JHEP}\ }\textbf {\bibinfo {volume} {05}},\ \bibinfo {pages} {055} (\bibinfo
  {year} {2020})},\ \Eprint {http://arxiv.org/abs/1912.00422} {arXiv:1912.00422
  [hep-ph]} \BibitemShut {NoStop}%
\bibitem [{\citenamefont {Aad}\ \emph {et~al.}(2016)\citenamefont {Aad} \emph
  {et~al.}}]{ATLAS:2016fij}%
  \BibitemOpen
  \bibfield  {author} {\bibinfo {author} {\bibfnamefont {G.}~\bibnamefont
  {Aad}} \emph {et~al.} (\bibinfo {collaboration} {ATLAS}),\ }\href {\doibase
  10.1016/j.physletb.2016.06.023} {\bibfield  {journal} {\bibinfo  {journal}
  {Phys. Lett. B}\ }\textbf {\bibinfo {volume} {759}},\ \bibinfo {pages} {601}
  (\bibinfo {year} {2016})},\ \Eprint {http://arxiv.org/abs/1603.09222}
  {arXiv:1603.09222 [hep-ex]} \BibitemShut {NoStop}%
\bibitem [{\citenamefont {Aad}\ \emph {et~al.}(2024{\natexlab{a}})\citenamefont
  {Aad} \emph {et~al.}}]{ATLAS:2024irg}%
  \BibitemOpen
  \bibfield  {author} {\bibinfo {author} {\bibfnamefont {G.}~\bibnamefont
  {Aad}} \emph {et~al.} (\bibinfo {collaboration} {ATLAS}),\ }\href {\doibase
  10.1016/j.physletb.2024.138725} {\bibfield  {journal} {\bibinfo  {journal}
  {Phys. Lett. B}\ }\textbf {\bibinfo {volume} {854}},\ \bibinfo {pages}
  {138725} (\bibinfo {year} {2024}{\natexlab{a}})},\ \Eprint
  {http://arxiv.org/abs/2403.12902} {arXiv:2403.12902 [hep-ex]} \BibitemShut
  {NoStop}%
\bibitem [{\citenamefont {Chekhovsky}\ \emph {et~al.}(2026)\citenamefont
  {Chekhovsky} \emph {et~al.}}]{CMS:2025sgv}%
  \BibitemOpen
  \bibfield  {author} {\bibinfo {author} {\bibfnamefont {V.}~\bibnamefont
  {Chekhovsky}} \emph {et~al.} (\bibinfo {collaboration} {CMS}),\ }\href
  {\doibase 10.1007/JHEP01(2026)047} {\bibfield  {journal} {\bibinfo  {journal}
  {JHEP}\ }\textbf {\bibinfo {volume} {01}},\ \bibinfo {pages} {047} (\bibinfo
  {year} {2026})},\ \Eprint {http://arxiv.org/abs/2503.09742} {arXiv:2503.09742
  [hep-ex]} \BibitemShut {NoStop}%
\bibitem [{\citenamefont {Sirunyan}\ \emph {et~al.}(2021)\citenamefont
  {Sirunyan} \emph {et~al.}}]{CMS:2021beq}%
  \BibitemOpen
  \bibfield  {author} {\bibinfo {author} {\bibfnamefont {A.~M.}\ \bibnamefont
  {Sirunyan}} \emph {et~al.} (\bibinfo {collaboration} {CMS}),\ }\href
  {\doibase 10.1103/PhysRevD.104.052001} {\bibfield  {journal} {\bibinfo
  {journal} {Phys. Rev. D}\ }\textbf {\bibinfo {volume} {104}},\ \bibinfo
  {pages} {052001} (\bibinfo {year} {2021})},\ \Eprint
  {http://arxiv.org/abs/2103.01290} {arXiv:2103.01290 [hep-ex]} \BibitemShut
  {NoStop}%
\bibitem [{\citenamefont {Aad}\ \emph {et~al.}(2024{\natexlab{b}})\citenamefont
  {Aad} \emph {et~al.}}]{ATLAS:2024rcx}%
  \BibitemOpen
  \bibfield  {author} {\bibinfo {author} {\bibfnamefont {G.}~\bibnamefont
  {Aad}} \emph {et~al.} (\bibinfo {collaboration} {ATLAS}),\ }\href {\doibase
  10.1007/JHEP03(2024)139} {\bibfield  {journal} {\bibinfo  {journal} {JHEP}\
  }\textbf {\bibinfo {volume} {03}},\ \bibinfo {pages} {139} (\bibinfo {year}
  {2024}{\natexlab{b}})},\ \Eprint {http://arxiv.org/abs/2401.13430}
  {arXiv:2401.13430 [hep-ex]} \BibitemShut {NoStop}%
\bibitem [{\citenamefont {Chekhovsky}\ \emph {et~al.}(2025)\citenamefont
  {Chekhovsky} \emph {et~al.}}]{CMS:2025ttk}%
  \BibitemOpen
  \bibfield  {author} {\bibinfo {author} {\bibfnamefont {V.}~\bibnamefont
  {Chekhovsky}} \emph {et~al.} (\bibinfo {collaboration} {CMS}),\ }\href@noop
  {} {\  (\bibinfo {year} {2025})},\ \Eprint {http://arxiv.org/abs/2508.13900}
  {arXiv:2508.13900 [hep-ex]} \BibitemShut {NoStop}%
\bibitem [{\citenamefont {Beenakker}\ \emph {et~al.}(2024)\citenamefont
  {Beenakker}, \citenamefont {Borschensky}, \citenamefont {Kr{\"a}mer},
  \citenamefont {Kulesza}, \citenamefont {Laenen}, \citenamefont
  {Mamu{\v{z}}i{\'c}},\ and\ \citenamefont {Valero}}]{Beenakker:2024jwh}%
  \BibitemOpen
  \bibfield  {author} {\bibinfo {author} {\bibfnamefont {W.}~\bibnamefont
  {Beenakker}}, \bibinfo {author} {\bibfnamefont {C.}~\bibnamefont
  {Borschensky}}, \bibinfo {author} {\bibfnamefont {M.}~\bibnamefont
  {Kr{\"a}mer}}, \bibinfo {author} {\bibfnamefont {A.}~\bibnamefont {Kulesza}},
  \bibinfo {author} {\bibfnamefont {E.}~\bibnamefont {Laenen}}, \bibinfo
  {author} {\bibfnamefont {J.}~\bibnamefont {Mamu{\v{z}}i{\'c}}}, \ and\
  \bibinfo {author} {\bibfnamefont {L.~M.}\ \bibnamefont {Valero}},\ }\href
  {\doibase 10.21468/SciPostPhysCore.7.4.072} {\bibfield  {journal} {\bibinfo
  {journal} {SciPost Phys. Core}\ }\textbf {\bibinfo {volume} {7}},\ \bibinfo
  {pages} {072} (\bibinfo {year} {2024})},\ \Eprint
  {http://arxiv.org/abs/2404.18837} {arXiv:2404.18837 [hep-ph]} \BibitemShut
  {NoStop}%
\bibitem [{\citenamefont {Sirunyan}\ \emph {et~al.}(2019)\citenamefont
  {Sirunyan} \emph {et~al.}}]{CMS:2018uag}%
  \BibitemOpen
  \bibfield  {author} {\bibinfo {author} {\bibfnamefont {A.~M.}\ \bibnamefont
  {Sirunyan}} \emph {et~al.} (\bibinfo {collaboration} {CMS}),\ }\href
  {\doibase 10.1140/epjc/s10052-019-6909-y} {\bibfield  {journal} {\bibinfo
  {journal} {Eur. Phys. J. C}\ }\textbf {\bibinfo {volume} {79}},\ \bibinfo
  {pages} {421} (\bibinfo {year} {2019})},\ \Eprint
  {http://arxiv.org/abs/1809.10733} {arXiv:1809.10733 [hep-ex]} \BibitemShut
  {NoStop}%
\bibitem [{\citenamefont {Aad}\ \emph {et~al.}(2022)\citenamefont {Aad} \emph
  {et~al.}}]{ATLAS:2022vkf}%
  \BibitemOpen
  \bibfield  {author} {\bibinfo {author} {\bibfnamefont {G.}~\bibnamefont
  {Aad}} \emph {et~al.} (\bibinfo {collaboration} {ATLAS}),\ }\href {\doibase
  10.1038/s41586-022-04893-w} {\bibfield  {journal} {\bibinfo  {journal}
  {Nature}\ }\textbf {\bibinfo {volume} {607}},\ \bibinfo {pages} {52}
  (\bibinfo {year} {2022})},\ \bibinfo {note} {[Erratum: Nature 612, E24
  (2022)]},\ \Eprint {http://arxiv.org/abs/2207.00092} {arXiv:2207.00092
  [hep-ex]} \BibitemShut {NoStop}%
\bibitem [{\citenamefont {Aad}\ \emph {et~al.}(2024{\natexlab{c}})\citenamefont
  {Aad} \emph {et~al.}}]{ATLAS:2023tnc}%
  \BibitemOpen
  \bibfield  {author} {\bibinfo {author} {\bibfnamefont {G.}~\bibnamefont
  {Aad}} \emph {et~al.} (\bibinfo {collaboration} {ATLAS}),\ }\href {\doibase
  10.1140/epjc/s10052-023-12130-5} {\bibfield  {journal} {\bibinfo  {journal}
  {Eur. Phys. J. C}\ }\textbf {\bibinfo {volume} {84}},\ \bibinfo {pages} {78}
  (\bibinfo {year} {2024}{\natexlab{c}})},\ \Eprint
  {http://arxiv.org/abs/2306.11379} {arXiv:2306.11379 [hep-ex]} \BibitemShut
  {NoStop}%
\bibitem [{\citenamefont {Liu}\ \emph {et~al.}(2018)\citenamefont {Liu},
  \citenamefont {Wang}, \citenamefont {Wang},\ and\ \citenamefont
  {Xue}}]{Liu:2017zdh}%
  \BibitemOpen
  \bibfield  {author} {\bibinfo {author} {\bibfnamefont {J.}~\bibnamefont
  {Liu}}, \bibinfo {author} {\bibfnamefont {L.-T.}\ \bibnamefont {Wang}},
  \bibinfo {author} {\bibfnamefont {X.-P.}\ \bibnamefont {Wang}}, \ and\
  \bibinfo {author} {\bibfnamefont {W.}~\bibnamefont {Xue}},\ }\href {\doibase
  10.1103/PhysRevD.97.095044} {\bibfield  {journal} {\bibinfo  {journal} {Phys.
  Rev. D}\ }\textbf {\bibinfo {volume} {97}},\ \bibinfo {pages} {095044}
  (\bibinfo {year} {2018})},\ \Eprint {http://arxiv.org/abs/1712.07237}
  {arXiv:1712.07237 [hep-ph]} \BibitemShut {NoStop}%
\bibitem [{\citenamefont {Cheng}\ \emph {et~al.}(2019)\citenamefont {Cheng},
  \citenamefont {Li}, \citenamefont {Salvioni},\ and\ \citenamefont
  {Verhaaren}}]{Cheng:2019yai}%
  \BibitemOpen
  \bibfield  {author} {\bibinfo {author} {\bibfnamefont {H.-C.}\ \bibnamefont
  {Cheng}}, \bibinfo {author} {\bibfnamefont {L.}~\bibnamefont {Li}}, \bibinfo
  {author} {\bibfnamefont {E.}~\bibnamefont {Salvioni}}, \ and\ \bibinfo
  {author} {\bibfnamefont {C.~B.}\ \bibnamefont {Verhaaren}},\ }\href {\doibase
  10.1007/JHEP11(2019)031} {\bibfield  {journal} {\bibinfo  {journal} {JHEP}\
  }\textbf {\bibinfo {volume} {11}},\ \bibinfo {pages} {031} (\bibinfo {year}
  {2019})},\ \Eprint {http://arxiv.org/abs/1906.02198} {arXiv:1906.02198
  [hep-ph]} \BibitemShut {NoStop}%
\bibitem [{\citenamefont {Cheng}\ \emph {et~al.}(2024)\citenamefont {Cheng},
  \citenamefont {Jiang}, \citenamefont {Li},\ and\ \citenamefont
  {Salvioni}}]{Cheng:2024hvq}%
  \BibitemOpen
  \bibfield  {author} {\bibinfo {author} {\bibfnamefont {H.-C.}\ \bibnamefont
  {Cheng}}, \bibinfo {author} {\bibfnamefont {X.-H.}\ \bibnamefont {Jiang}},
  \bibinfo {author} {\bibfnamefont {L.}~\bibnamefont {Li}}, \ and\ \bibinfo
  {author} {\bibfnamefont {E.}~\bibnamefont {Salvioni}},\ }\href {\doibase
  10.1007/JHEP04(2024)081} {\bibfield  {journal} {\bibinfo  {journal} {JHEP}\
  }\textbf {\bibinfo {volume} {04}},\ \bibinfo {pages} {081} (\bibinfo {year}
  {2024})},\ \Eprint {http://arxiv.org/abs/2401.08785} {arXiv:2401.08785
  [hep-ph]} \BibitemShut {NoStop}%
\bibitem [{\citenamefont {Schael}\ \emph {et~al.}(2006)\citenamefont {Schael}
  \emph {et~al.}}]{ALEPH:2005ab}%
  \BibitemOpen
  \bibfield  {author} {\bibinfo {author} {\bibfnamefont {S.}~\bibnamefont
  {Schael}} \emph {et~al.} (\bibinfo {collaboration} {ALEPH, DELPHI, L3, OPAL,
  SLD, LEP Electroweak Working Group, SLD Electroweak Group, SLD Heavy Flavour
  Group}),\ }\href {\doibase 10.1016/j.physrep.2005.12.006} {\bibfield
  {journal} {\bibinfo  {journal} {Phys. Rept.}\ }\textbf {\bibinfo {volume}
  {427}},\ \bibinfo {pages} {257} (\bibinfo {year} {2006})},\ \Eprint
  {http://arxiv.org/abs/hep-ex/0509008} {arXiv:hep-ex/0509008} \BibitemShut
  {NoStop}%
\bibitem [{\citenamefont {Navas}\ \emph {et~al.}(2024)\citenamefont {Navas}
  \emph {et~al.}}]{ParticleDataGroup:2024cfk}%
  \BibitemOpen
  \bibfield  {author} {\bibinfo {author} {\bibfnamefont {S.}~\bibnamefont
  {Navas}} \emph {et~al.} (\bibinfo {collaboration} {Particle Data Group}),\
  }\href {\doibase 10.1103/PhysRevD.110.030001} {\bibfield  {journal} {\bibinfo
   {journal} {Phys. Rev. D}\ }\textbf {\bibinfo {volume} {110}},\ \bibinfo
  {pages} {030001} (\bibinfo {year} {2024})}\BibitemShut {NoStop}%
\bibitem [{\citenamefont {Abdallah}\ \emph {et~al.}(2015)\citenamefont
  {Abdallah} \emph {et~al.}}]{Abdallah:2015ter}%
  \BibitemOpen
  \bibfield  {author} {\bibinfo {author} {\bibfnamefont {J.}~\bibnamefont
  {Abdallah}} \emph {et~al.},\ }\href {\doibase 10.1016/j.dark.2015.08.001}
  {\bibfield  {journal} {\bibinfo  {journal} {Phys. Dark Univ.}\ }\textbf
  {\bibinfo {volume} {9-10}},\ \bibinfo {pages} {8} (\bibinfo {year} {2015})},\
  \Eprint {http://arxiv.org/abs/1506.03116} {arXiv:1506.03116 [hep-ph]}
  \BibitemShut {NoStop}%
\bibitem [{\citenamefont {Olive}\ \emph {et~al.}(2014)\citenamefont {Olive}
  \emph {et~al.}}]{ParticleDataGroup:2014cgo}%
  \BibitemOpen
  \bibfield  {author} {\bibinfo {author} {\bibfnamefont {K.~A.}\ \bibnamefont
  {Olive}} \emph {et~al.} (\bibinfo {collaboration} {Particle Data Group}),\
  }\href {\doibase 10.1088/1674-1137/38/9/090001} {\bibfield  {journal}
  {\bibinfo  {journal} {Chin. Phys. C}\ }\textbf {\bibinfo {volume} {38}},\
  \bibinfo {pages} {090001} (\bibinfo {year} {2014})}\BibitemShut {NoStop}%
\bibitem [{\citenamefont {Feng}\ \emph {et~al.}(2011)\citenamefont {Feng},
  \citenamefont {Kumar}, \citenamefont {Marfatia},\ and\ \citenamefont
  {Sanford}}]{Feng:2011vu}%
  \BibitemOpen
  \bibfield  {author} {\bibinfo {author} {\bibfnamefont {J.~L.}\ \bibnamefont
  {Feng}}, \bibinfo {author} {\bibfnamefont {J.}~\bibnamefont {Kumar}},
  \bibinfo {author} {\bibfnamefont {D.}~\bibnamefont {Marfatia}}, \ and\
  \bibinfo {author} {\bibfnamefont {D.}~\bibnamefont {Sanford}},\ }\href
  {\doibase 10.1016/j.physletb.2011.07.083} {\bibfield  {journal} {\bibinfo
  {journal} {Phys. Lett. B}\ }\textbf {\bibinfo {volume} {703}},\ \bibinfo
  {pages} {124} (\bibinfo {year} {2011})},\ \Eprint
  {http://arxiv.org/abs/1102.4331} {arXiv:1102.4331 [hep-ph]} \BibitemShut
  {NoStop}%
\bibitem [{\citenamefont {Aalbers}\ \emph {et~al.}(2025)\citenamefont {Aalbers}
  \emph {et~al.}}]{LZ:2024zvo}%
  \BibitemOpen
  \bibfield  {author} {\bibinfo {author} {\bibfnamefont {J.}~\bibnamefont
  {Aalbers}} \emph {et~al.} (\bibinfo {collaboration} {LZ}),\ }\href {\doibase
  10.1103/4dyc-z8zf} {\bibfield  {journal} {\bibinfo  {journal} {Phys. Rev.
  Lett.}\ }\textbf {\bibinfo {volume} {135}},\ \bibinfo {pages} {011802}
  (\bibinfo {year} {2025})},\ \Eprint {http://arxiv.org/abs/2410.17036}
  {arXiv:2410.17036 [hep-ex]} \BibitemShut {NoStop}%
\bibitem [{\citenamefont {Aprile}\ \emph {et~al.}(2019)\citenamefont {Aprile}
  \emph {et~al.}}]{XENON:2019gfn}%
  \BibitemOpen
  \bibfield  {author} {\bibinfo {author} {\bibfnamefont {E.}~\bibnamefont
  {Aprile}} \emph {et~al.} (\bibinfo {collaboration} {XENON}),\ }\href
  {\doibase 10.1103/PhysRevLett.123.251801} {\bibfield  {journal} {\bibinfo
  {journal} {Phys. Rev. Lett.}\ }\textbf {\bibinfo {volume} {123}},\ \bibinfo
  {pages} {251801} (\bibinfo {year} {2019})},\ \Eprint
  {http://arxiv.org/abs/1907.11485} {arXiv:1907.11485 [hep-ex]} \BibitemShut
  {NoStop}%
\bibitem [{\citenamefont {Agnes}\ \emph {et~al.}(2023)\citenamefont {Agnes}
  \emph {et~al.}}]{DarkSide-50:2022qzh}%
  \BibitemOpen
  \bibfield  {author} {\bibinfo {author} {\bibfnamefont {P.}~\bibnamefont
  {Agnes}} \emph {et~al.} (\bibinfo {collaboration} {DarkSide-50}),\ }\href
  {\doibase 10.1103/PhysRevD.107.063001} {\bibfield  {journal} {\bibinfo
  {journal} {Phys. Rev. D}\ }\textbf {\bibinfo {volume} {107}},\ \bibinfo
  {pages} {063001} (\bibinfo {year} {2023})},\ \Eprint
  {http://arxiv.org/abs/2207.11966} {arXiv:2207.11966 [hep-ex]} \BibitemShut
  {NoStop}%
\bibitem [{\citenamefont {Acerbi}\ \emph {et~al.}(2025)\citenamefont {Acerbi}
  \emph {et~al.}}]{DarkSide-50:2025lns}%
  \BibitemOpen
  \bibfield  {author} {\bibinfo {author} {\bibfnamefont {F.}~\bibnamefont
  {Acerbi}} \emph {et~al.} (\bibinfo {collaboration} {DarkSide-50,
  DarkSide-20k}),\ }\href@noop {} {\  (\bibinfo {year} {2025})},\ \Eprint
  {http://arxiv.org/abs/2511.13629} {arXiv:2511.13629 [hep-ex]} \BibitemShut
  {NoStop}%
\bibitem [{\citenamefont {Amole}\ \emph {et~al.}(2019)\citenamefont {Amole}
  \emph {et~al.}}]{PICO:2019vsc}%
  \BibitemOpen
  \bibfield  {author} {\bibinfo {author} {\bibfnamefont {C.}~\bibnamefont
  {Amole}} \emph {et~al.} (\bibinfo {collaboration} {PICO}),\ }\href {\doibase
  10.1103/PhysRevD.100.022001} {\bibfield  {journal} {\bibinfo  {journal}
  {Phys. Rev. D}\ }\textbf {\bibinfo {volume} {100}},\ \bibinfo {pages}
  {022001} (\bibinfo {year} {2019})},\ \Eprint
  {http://arxiv.org/abs/1902.04031} {arXiv:1902.04031 [astro-ph.CO]}
  \BibitemShut {NoStop}%
\bibitem [{\citenamefont {Tumasyan}\ \emph {et~al.}(2021)\citenamefont
  {Tumasyan} \emph {et~al.}}]{CMS:2021far}%
  \BibitemOpen
  \bibfield  {author} {\bibinfo {author} {\bibfnamefont {A.}~\bibnamefont
  {Tumasyan}} \emph {et~al.} (\bibinfo {collaboration} {CMS}),\ }\href
  {\doibase 10.1007/JHEP11(2021)153} {\bibfield  {journal} {\bibinfo  {journal}
  {JHEP}\ }\textbf {\bibinfo {volume} {11}},\ \bibinfo {pages} {153} (\bibinfo
  {year} {2021})},\ \Eprint {http://arxiv.org/abs/2107.13021} {arXiv:2107.13021
  [hep-ex]} \BibitemShut {NoStop}%
\bibitem [{\citenamefont {Aad}\ \emph {et~al.}(2021)\citenamefont {Aad} \emph
  {et~al.}}]{ATLAS:2021kxv}%
  \BibitemOpen
  \bibfield  {author} {\bibinfo {author} {\bibfnamefont {G.}~\bibnamefont
  {Aad}} \emph {et~al.} (\bibinfo {collaboration} {ATLAS}),\ }\href {\doibase
  10.1103/PhysRevD.103.112006} {\bibfield  {journal} {\bibinfo  {journal}
  {Phys. Rev. D}\ }\textbf {\bibinfo {volume} {103}},\ \bibinfo {pages}
  {112006} (\bibinfo {year} {2021})},\ \Eprint
  {http://arxiv.org/abs/2102.10874} {arXiv:2102.10874 [hep-ex]} \BibitemShut
  {NoStop}%
\bibitem [{\citenamefont {Aghanim}\ \emph {et~al.}(2020)\citenamefont {Aghanim}
  \emph {et~al.}}]{Planck:2018vyg}%
  \BibitemOpen
  \bibfield  {author} {\bibinfo {author} {\bibfnamefont {N.}~\bibnamefont
  {Aghanim}} \emph {et~al.} (\bibinfo {collaboration} {Planck}),\ }\href
  {\doibase 10.1051/0004-6361/201833910} {\bibfield  {journal} {\bibinfo
  {journal} {Astron. Astrophys.}\ }\textbf {\bibinfo {volume} {641}},\ \bibinfo
  {pages} {A6} (\bibinfo {year} {2020})},\ \bibinfo {note} {[Erratum:
  Astron.Astrophys. 652, C4 (2021)]},\ \Eprint
  {http://arxiv.org/abs/1807.06209} {arXiv:1807.06209 [astro-ph.CO]}
  \BibitemShut {NoStop}%
\bibitem [{\citenamefont {Alguero}\ \emph {et~al.}(2024)\citenamefont
  {Alguero}, \citenamefont {Belanger}, \citenamefont {Boudjema}, \citenamefont
  {Chakraborti}, \citenamefont {Goudelis}, \citenamefont {Kraml}, \citenamefont
  {Mjallal},\ and\ \citenamefont {Pukhov}}]{Alguero:2023zol}%
  \BibitemOpen
  \bibfield  {author} {\bibinfo {author} {\bibfnamefont {G.}~\bibnamefont
  {Alguero}}, \bibinfo {author} {\bibfnamefont {G.}~\bibnamefont {Belanger}},
  \bibinfo {author} {\bibfnamefont {F.}~\bibnamefont {Boudjema}}, \bibinfo
  {author} {\bibfnamefont {S.}~\bibnamefont {Chakraborti}}, \bibinfo {author}
  {\bibfnamefont {A.}~\bibnamefont {Goudelis}}, \bibinfo {author}
  {\bibfnamefont {S.}~\bibnamefont {Kraml}}, \bibinfo {author} {\bibfnamefont
  {A.}~\bibnamefont {Mjallal}}, \ and\ \bibinfo {author} {\bibfnamefont
  {A.}~\bibnamefont {Pukhov}},\ }\href {\doibase 10.1016/j.cpc.2024.109133}
  {\bibfield  {journal} {\bibinfo  {journal} {Comput. Phys. Commun.}\ }\textbf
  {\bibinfo {volume} {299}},\ \bibinfo {pages} {109133} (\bibinfo {year}
  {2024})},\ \Eprint {http://arxiv.org/abs/2312.14894} {arXiv:2312.14894
  [hep-ph]} \BibitemShut {NoStop}%
\bibitem [{\citenamefont {Feng}\ \emph {et~al.}(2018)\citenamefont {Feng},
  \citenamefont {Galon}, \citenamefont {Kling},\ and\ \citenamefont
  {Trojanowski}}]{Feng:2017uoz}%
  \BibitemOpen
  \bibfield  {author} {\bibinfo {author} {\bibfnamefont {J.~L.}\ \bibnamefont
  {Feng}}, \bibinfo {author} {\bibfnamefont {I.}~\bibnamefont {Galon}},
  \bibinfo {author} {\bibfnamefont {F.}~\bibnamefont {Kling}}, \ and\ \bibinfo
  {author} {\bibfnamefont {S.}~\bibnamefont {Trojanowski}},\ }\href {\doibase
  10.1103/PhysRevD.97.035001} {\bibfield  {journal} {\bibinfo  {journal} {Phys.
  Rev. D}\ }\textbf {\bibinfo {volume} {97}},\ \bibinfo {pages} {035001}
  (\bibinfo {year} {2018})},\ \Eprint {http://arxiv.org/abs/1708.09389}
  {arXiv:1708.09389 [hep-ph]} \BibitemShut {NoStop}%
\bibitem [{\citenamefont {Ariga}\ \emph {et~al.}(2019)\citenamefont {Ariga}
  \emph {et~al.}}]{FASER:2018eoc}%
  \BibitemOpen
  \bibfield  {author} {\bibinfo {author} {\bibfnamefont {A.}~\bibnamefont
  {Ariga}} \emph {et~al.} (\bibinfo {collaboration} {FASER}),\ }\href {\doibase
  10.1103/PhysRevD.99.095011} {\bibfield  {journal} {\bibinfo  {journal} {Phys.
  Rev. D}\ }\textbf {\bibinfo {volume} {99}},\ \bibinfo {pages} {095011}
  (\bibinfo {year} {2019})},\ \Eprint {http://arxiv.org/abs/1811.12522}
  {arXiv:1811.12522 [hep-ph]} \BibitemShut {NoStop}%
\bibitem [{\citenamefont {Cerci}\ \emph {et~al.}(2022)\citenamefont {Cerci}
  \emph {et~al.}}]{Cerci:2021nlb}%
  \BibitemOpen
  \bibfield  {author} {\bibinfo {author} {\bibfnamefont {S.}~\bibnamefont
  {Cerci}} \emph {et~al.},\ }\href {\doibase 10.1007/JHEP06(2022)110}
  {\bibfield  {journal} {\bibinfo  {journal} {JHEP}\ }\textbf {\bibinfo
  {volume} {06}},\ \bibinfo {pages} {110} (\bibinfo {year} {2022})},\ \Eprint
  {http://arxiv.org/abs/2201.00019} {arXiv:2201.00019 [hep-ex]} \BibitemShut
  {NoStop}%
\bibitem [{\citenamefont {Aitken}\ \emph {et~al.}(2025)\citenamefont {Aitken}
  \emph {et~al.}}]{MATHUSLA:2025zyt}%
  \BibitemOpen
  \bibfield  {author} {\bibinfo {author} {\bibfnamefont {B.}~\bibnamefont
  {Aitken}} \emph {et~al.} (\bibinfo {collaboration} {MATHUSLA}),\ }\href@noop
  {} {\  (\bibinfo {year} {2025})},\ \Eprint {http://arxiv.org/abs/2503.20893}
  {arXiv:2503.20893 [physics.ins-det]} \BibitemShut {NoStop}%
\bibitem [{\citenamefont {Alwall}\ \emph {et~al.}(2014)\citenamefont {Alwall},
  \citenamefont {Frederix}, \citenamefont {Frixione}, \citenamefont {Hirschi},
  \citenamefont {Maltoni}, \citenamefont {Mattelaer}, \citenamefont {Shao},
  \citenamefont {Stelzer}, \citenamefont {Torrielli},\ and\ \citenamefont
  {Zaro}}]{Alwall:2014hca}%
  \BibitemOpen
  \bibfield  {author} {\bibinfo {author} {\bibfnamefont {J.}~\bibnamefont
  {Alwall}}, \bibinfo {author} {\bibfnamefont {R.}~\bibnamefont {Frederix}},
  \bibinfo {author} {\bibfnamefont {S.}~\bibnamefont {Frixione}}, \bibinfo
  {author} {\bibfnamefont {V.}~\bibnamefont {Hirschi}}, \bibinfo {author}
  {\bibfnamefont {F.}~\bibnamefont {Maltoni}}, \bibinfo {author} {\bibfnamefont
  {O.}~\bibnamefont {Mattelaer}}, \bibinfo {author} {\bibfnamefont {H.~S.}\
  \bibnamefont {Shao}}, \bibinfo {author} {\bibfnamefont {T.}~\bibnamefont
  {Stelzer}}, \bibinfo {author} {\bibfnamefont {P.}~\bibnamefont {Torrielli}},
  \ and\ \bibinfo {author} {\bibfnamefont {M.}~\bibnamefont {Zaro}},\ }\href
  {\doibase 10.1007/JHEP07(2014)079} {\bibfield  {journal} {\bibinfo  {journal}
  {JHEP}\ }\textbf {\bibinfo {volume} {07}},\ \bibinfo {pages} {079} (\bibinfo
  {year} {2014})},\ \Eprint {http://arxiv.org/abs/1405.0301} {arXiv:1405.0301
  [hep-ph]} \BibitemShut {NoStop}%
\bibitem [{\citenamefont {Sj{\"o}strand}\ \emph {et~al.}(2015)\citenamefont
  {Sj{\"o}strand}, \citenamefont {Ask}, \citenamefont {Christiansen},
  \citenamefont {Corke}, \citenamefont {Desai}, \citenamefont {Ilten},
  \citenamefont {Mrenna}, \citenamefont {Prestel}, \citenamefont {Rasmussen},\
  and\ \citenamefont {Skands}}]{Sjostrand:2014zea}%
  \BibitemOpen
  \bibfield  {author} {\bibinfo {author} {\bibfnamefont {T.}~\bibnamefont
  {Sj{\"o}strand}}, \bibinfo {author} {\bibfnamefont {S.}~\bibnamefont {Ask}},
  \bibinfo {author} {\bibfnamefont {J.~R.}\ \bibnamefont {Christiansen}},
  \bibinfo {author} {\bibfnamefont {R.}~\bibnamefont {Corke}}, \bibinfo
  {author} {\bibfnamefont {N.}~\bibnamefont {Desai}}, \bibinfo {author}
  {\bibfnamefont {P.}~\bibnamefont {Ilten}}, \bibinfo {author} {\bibfnamefont
  {S.}~\bibnamefont {Mrenna}}, \bibinfo {author} {\bibfnamefont
  {S.}~\bibnamefont {Prestel}}, \bibinfo {author} {\bibfnamefont {C.~O.}\
  \bibnamefont {Rasmussen}}, \ and\ \bibinfo {author} {\bibfnamefont {P.~Z.}\
  \bibnamefont {Skands}},\ }\href {\doibase 10.1016/j.cpc.2015.01.024}
  {\bibfield  {journal} {\bibinfo  {journal} {Comput. Phys. Commun.}\ }\textbf
  {\bibinfo {volume} {191}},\ \bibinfo {pages} {159} (\bibinfo {year}
  {2015})},\ \Eprint {http://arxiv.org/abs/1410.3012} {arXiv:1410.3012
  [hep-ph]} \BibitemShut {NoStop}%
\bibitem [{\citenamefont {Kling}\ and\ \citenamefont
  {Trojanowski}(2021)}]{Kling:2021fwx}%
  \BibitemOpen
  \bibfield  {author} {\bibinfo {author} {\bibfnamefont {F.}~\bibnamefont
  {Kling}}\ and\ \bibinfo {author} {\bibfnamefont {S.}~\bibnamefont
  {Trojanowski}},\ }\href {\doibase 10.1103/PhysRevD.104.035012} {\bibfield
  {journal} {\bibinfo  {journal} {Phys. Rev. D}\ }\textbf {\bibinfo {volume}
  {104}},\ \bibinfo {pages} {035012} (\bibinfo {year} {2021})},\ \Eprint
  {http://arxiv.org/abs/2105.07077} {arXiv:2105.07077 [hep-ph]} \BibitemShut
  {NoStop}%
\bibitem [{\citenamefont {Batell}\ \emph {et~al.}(2012)\citenamefont {Batell},
  \citenamefont {Gori},\ and\ \citenamefont {Wang}}]{Batell:2011pz}%
  \BibitemOpen
  \bibfield  {author} {\bibinfo {author} {\bibfnamefont {B.}~\bibnamefont
  {Batell}}, \bibinfo {author} {\bibfnamefont {S.}~\bibnamefont {Gori}}, \ and\
  \bibinfo {author} {\bibfnamefont {L.-T.}\ \bibnamefont {Wang}},\ }\href
  {\doibase 10.1007/JHEP06(2012)172} {\bibfield  {journal} {\bibinfo  {journal}
  {JHEP}\ }\textbf {\bibinfo {volume} {06}},\ \bibinfo {pages} {172} (\bibinfo
  {year} {2012})},\ \Eprint {http://arxiv.org/abs/1112.5180} {arXiv:1112.5180
  [hep-ph]} \BibitemShut {NoStop}%
\bibitem [{\citenamefont {Kribs}\ and\ \citenamefont
  {Martin}(2012)}]{Kribs:2012kz}%
  \BibitemOpen
  \bibfield  {author} {\bibinfo {author} {\bibfnamefont {G.~D.}\ \bibnamefont
  {Kribs}}\ and\ \bibinfo {author} {\bibfnamefont {A.}~\bibnamefont {Martin}},\
  }\href {\doibase 10.1103/PhysRevD.86.095023} {\bibfield  {journal} {\bibinfo
  {journal} {Phys. Rev. D}\ }\textbf {\bibinfo {volume} {86}},\ \bibinfo
  {pages} {095023} (\bibinfo {year} {2012})},\ \Eprint
  {http://arxiv.org/abs/1207.4496} {arXiv:1207.4496 [hep-ph]} \BibitemShut
  {NoStop}%
\bibitem [{\citenamefont {Passarino}\ and\ \citenamefont
  {Veltman}(1979)}]{Passarino:1978jh}%
  \BibitemOpen
  \bibfield  {author} {\bibinfo {author} {\bibfnamefont {G.}~\bibnamefont
  {Passarino}}\ and\ \bibinfo {author} {\bibfnamefont {M.~J.~G.}\ \bibnamefont
  {Veltman}},\ }\href {\doibase 10.1016/0550-3213(79)90234-7} {\bibfield
  {journal} {\bibinfo  {journal} {Nucl. Phys. B}\ }\textbf {\bibinfo {volume}
  {160}},\ \bibinfo {pages} {151} (\bibinfo {year} {1979})}\BibitemShut
  {NoStop}%
\end{thebibliography}%

\end{document}